\documentclass[twocolumn,astrosymb]{aastex62}

\usepackage{morefloats}
\usepackage{amsmath}
\usepackage{amssymb}
\usepackage{latexsym}
\usepackage{multirow}
\usepackage{enumitem}
\hypersetup{linkcolor=red,citecolor=blue,filecolor=cyan,urlcolor=teal}
\received{\today}
\revised{N/A}
\accepted{N/A}
\submitjournal{ApJS}
\shorttitle{SOLAR WIND STATS}
\shortauthors{Wilson III et al.}

\begin{document}

\title{The statistical properties of solar wind temperature parameters near 1 AU}
\correspondingauthor{L.B. Wilson III}
\email{lynn.b.wilsoniii@gmail.com}

\author[0000-0002-4313-1970]{Lynn B. Wilson III}
\affiliation{NASA Goddard Space Flight Center, Heliophysics Science Division, Greenbelt, Maryland, USA.}

\author[0000-0002-7728-0085]{Michael L. Stevens}
\affiliation{Harvard-Smithsonian Center for Astrophysics, Harvard University, Cambridge, Massachusetts, USA.}

\author[0000-0002-7077-930X]{Justin C. Kasper}
\affiliation{University of Michigan, Ann Arbor, School of Climate and Space Sciences and Engineering, Ann Arbor, Michigan, USA.}

\author[0000-0001-6038-1923]{Kristopher G. Klein}
\affiliation{University of Michigan, Ann Arbor, School of Climate and Space Sciences and Engineering, Ann Arbor, Michigan, USA.}
\affiliation{Lunar and Planetary Laboratory, University of Arizona, Tucson, AZ 85719, USA.}

\author[0000-0002-2229-5618]{Bennett A. Maruca}
\affiliation{University of Delaware, Department of Physics \& Astronomy, Newark, Delaware, USA.}

\author[0000-0002-1989-3596]{Stuart D. Bale}
\affiliation{University of California Berkeley, Space Sciences Laboratory, Berkeley, California, USA.}

\author[0000-0002-4625-3332]{Trevor A. Bowen}
\affiliation{University of California Berkeley, Space Sciences Laboratory, Berkeley, California, USA.}

\author[0000-0002-1573-7457]{Marc P. Pulupa}
\affiliation{University of California Berkeley, Space Sciences Laboratory, Berkeley, California, USA.}

\author[0000-0002-6536-1531]{Chadi S. Salem}
\affiliation{University of California Berkeley, Space Sciences Laboratory, Berkeley, California, USA.}

\begin{abstract}
  We present a long-duration ($\sim$10 years) statistical analysis of the temperatures, plasma betas, and temperature ratios for the electron, proton, and alpha-particle populations observed by the \emph{Wind} spacecraft near 1 AU.  The mean(median) scalar temperatures are $T{\scriptstyle_{e, tot}}$ $=$ 12.2(11.9) eV, $T{\scriptstyle_{p, tot}}$ $=$ 12.7(8.6) eV, and $T{\scriptstyle_{\alpha, tot}}$ $=$ 23.9(10.8) eV.  The mean(median) total plasma betas are $\beta{\scriptstyle_{e, tot}}$ $=$ 2.31(1.09), $\beta{\scriptstyle_{p, tot}}$ $=$ 1.79(1.05), and $\beta{\scriptstyle_{\alpha, tot}}$ $=$ 0.17(0.05).  The mean(median) temperature ratios are $\left(T{\scriptstyle_{e}}/T{\scriptstyle_{p}}\right){\scriptstyle_{tot}}$ $=$ 1.64(1.27), $\left(T{\scriptstyle_{e}}/T{\scriptstyle_{\alpha}}\right){\scriptstyle_{tot}}$ $=$ 1.24(0.82), and $\left(T{\scriptstyle_{\alpha}}/T{\scriptstyle_{p}}\right){\scriptstyle_{tot}}$ $=$ 2.50(1.94).  We also examined these parameters during time intervals that exclude interplanetary (IP) shocks, times within the magnetic obstacles (MOs) of interplanetary coronal mass ejections (ICMEs), and times that exclude MOs.  The only times that show significant alterations to any of the parameters examined are those during MOs.  In fact, the only parameter that does not show a significant change during MOs is the electron temperature.  Although each parameter shows a broad range of values, the vast majority are near the median.  We also compute particle-particle collision rates and compare to effective wave-particle collision rates.  We find that, for reasonable assumptions of wave amplitude and occurrence rates, the effect of wave-particle interactions on the plasma is equal to or greater than the effect of Coulomb collisions.  Thus, wave-particle interactions should not be neglected when modeling the solar wind.
\end{abstract}


\phantomsection   
\section{Background and Motivation}  \label{sec:introduction}

\indent  Understanding the relationship between various macroscopic parameters for the different species of a gas is critical for understanding the evolution and dynamics of said gas.  A gas in thermodynamic equilibrium exhibits equal temperatures between all constituent species, i.e., $\left(T{\scriptstyle_{s'}}/T{\scriptstyle_{s}}\right){\scriptstyle_{tot}}$ $=$ 1 for $s'$ $\neq$ $s$ (see Appendix \ref{app:Definitions} for further details and parameter/symbol definitions) and does not allow for heat flow.  The phase-space distributions for the constituents of a gas in thermodynamic equilibrium are isotropic, they exhibit no skewness (i.e., heat flux), and they are centered at the same bulk flow velocity.  A subtle contrast exists for thermal equilibrium where one still maintains $\left(T{\scriptstyle_{s'}}/T{\scriptstyle_{s}}\right){\scriptstyle_{tot}}$ $=$ 1 for $s'$ $\neq$ $s$ but this does not require isotropic or uniformly flowing velocity distributions, e.g., one can have heat fluxes or counter-streaming populations \citep[e.g.,][]{evans90a, hoover86a}.  A non-equilibrium gas can exhibit $\left(T{\scriptstyle_{s'}}/T{\scriptstyle_{s}}\right){\scriptstyle_{tot}}$ $\neq$ 1, among other departures from a maximal entropy state.  If the temperatures are mass-proportional, i.e., uniform thermal speeds, then the species can be said to have the same velocity distribution \citep[e.g.,][]{ogilvie69a}.

\indent  Generally, a gas requires some form of irreversible energy dissipation and transfer between species to reach thermodynamic equilibrium.  In the Earth's atmosphere, the primary mechanism is binary particle collisions \citep[e.g.,][]{petschek58, shu92a}.  In the ionized gas or plasma of the solar wind Coulomb collisions play a role but the collision rate is often too low to drive $\left(T{\scriptstyle_{s'}}/T{\scriptstyle_{s}}\right){\scriptstyle_{tot}}$ $\rightarrow$ 1 as the plasma convects to Earth from the sun \citep[e.g.,][]{kasper08a, maruca13b}.  Because the collision rate is often so low, the solar wind is typically considered a collisionless plasma (or at the very least a weakly collisional plasma).  Determining the initial non-equilibrium $\left(T{\scriptstyle_{s'}}/T{\scriptstyle_{s}}\right){\scriptstyle_{tot}}$ $\neq$ 1 state of the gas and how it is established is critical to understanding its evolution and dynamics.

\indent  The relationship between various plasma parameters for the different major solar wind species is not well understood but understanding is critical for predicting the evolution and dynamics of the solar wind.  Some of the most important parameters are (see Appendix \ref{app:Definitions} for definitions):  the scalar temperatures $T{\scriptstyle_{s, j}}$ of species $s$, temperature ratios $\left(T{\scriptstyle_{s'}}/T{\scriptstyle_{s}}\right){\scriptstyle_{j}}$, and plasma betas $\beta{\scriptstyle_{s, j}}$ (where $j$ $=$ $tot$, $\parallel$, or $\perp$, i.e., the component with respect to the quasi-static magnetic field, $\textbf{\textit{B}}{\scriptstyle_{o}}$).  While some of these parameters have been measured in previous work (e.g., see Tables \ref{tab:PrevTemperatures}, \ref{tab:PrevTempRatios}, and \ref{tab:PrevBetas} in Appendix \ref{app:PreviousMeasurements}), we are not aware of a comprehensive statistical study of these parameters in the literature using data from a single spacecraft.

\indent  Many of these plasma parameters are critically important for theoretical and observational work using spacecraft that cannot measure, for instance, the electron distributions.  Most previous studies were case studies or limited to one parameter without summarizing tables providing quantities for future reference and independent of comparison to other parameters.  For instance, the study by \citet[][]{newbury98a} is one of the only studies that directly compared the electron and proton temperatures for a long-duration (i.e., more than 1 year), statistically significant dataset (they used 18 months of ISEE 3 data).  It is also one of the most often cited works\footnote{There were 71 citations on SAO/NASA ADS as of~\today.} for the average values of the ratio $\left(T{\scriptstyle_{e}}/T{\scriptstyle_{p}}\right){\scriptstyle_{tot}}$ in the solar wind.  However, it relied on five minute averages for only 18 months of data or $\sim$160,000 measurements.

\indent  One of the reasons there have been no long-duration statistical studies is because making an accurate measurement of, for instance, the full electron velocity distribution is very difficult.  The spacecraft needs to unambiguously measure the total electron density using quasi-thermal noise spectroscopy \citep[e.g.,][]{meyervernet89a, pulupa14a, pulupa14b} or the spacecraft electric potential or both.  Both of these measurements require accurate electric field instrumentation with the former requiring radio frequency measurements above the local upper hybrid frequency \citep[e.g.,][]{pulupa14a, pulupa14b} and the latter accurate, DC-coupled monopole measurements \citep[e.g.,][]{cully07a}.  Only recently with the high quality instrumentation and calibrated measurements by the \emph{Wind} spacecraft have truly long-duration, single spacecraft statistical studies of the solar wind been possible.

\indent  In this paper we describe the first long-duration statistical analysis of the temperatures, plasma betas, and temperature ratios of electrons, protons, and alpha-particles observed by the \emph{Wind} spacecraft near 1 AU between January 1995 and January 2005 comprised of over one million measurements.  That is, this study spans from the end of solar cycle 22 (i.e., March 1986 to May 1996) through much of solar cycle 23 (i.e., June 1996 to December 2008).  However, since the study does not span multiple complete solar cycles, we cannot perform a proper solar cycle dependence analysis.  This work is timely as it will provide a statistical baseline for the upcoming \emph{Solar Orbiter} \citep[][]{muller13a} and \emph{Parker Solar Probe} \citep[][]{fox16a} and future IMAP missions.

\indent  The paper is outlined as follows:  Section \ref{sec:DefinitionsDataSets} introduces the data sets, databases, and methodology used herein; Section \ref{sec:SolarWindStatistics} describes the statistical results; and Section \ref{sec:DiscussionandConclusions} presents the discussion and conclusion.  We also include appendices that provide additional details for the reader of the parameter definitions (Appendix \ref{app:Definitions}), collision rates (Appendix \ref{app:CoulombCollisions}), and summaries of previous work (Appendix \ref{app:PreviousMeasurements}).

\section{Data Sets and Methodology}  \label{sec:DefinitionsDataSets}

\indent  In this section we introduce the instrument data sets, shock database, interplanetary coronal mass ejection (ICME) catalog, the data selection method, and the analysis techniques used to examine the solar wind plasma observed by the \emph{Wind} spacecraft \citep{harten95a} near 1 AU.  The symbol/parameter definitions can be found in Appendix \ref{app:Definitions}.  Note that the purpose of the study is not to evaluate the quality of the datasets but to provide a concise summary of the statistical properties of some critically important solar wind parameters near 1 AU.

\subsection{Instruments}  \label{subsec:Instruments}

\indent  Quasi-static magnetic field vectors ($\textbf{\textit{B}}{\scriptstyle_{o}}$) were taken from the \emph{Wind}/MFI dual, triaxial fluxgate magnetometers \citep[][]{lepping95} using the one minute cadence data.  The components of some parameters used herein are defined with respect to $\textbf{\textit{B}}{\scriptstyle_{o}}$ using the subscript $j$.  That is, for all temperature-dependent parameters we compute the values for the entire distribution ($j$ $=$ $tot$), the parallel component ($j$ $=$ $\parallel$), and the perpendicular component ($j$ $=$ $\perp$).

\indent  The proton and alpha-particle densities ($n{\scriptstyle_{s}}$) and temperatures ($T{\scriptstyle_{s, j}}$) were taken from the \emph{Wind}/SWE Faraday Cups (FCs) \citep[][]{ogilvie95} at a $\sim$92 second cadence covering an energy-per-charge range of $\sim$150--8000 eV/C.  The velocity moments were calculated using a nonlinear least-squares fit of bi-Maxwellians to each species \citep[][]{kasper06a}.  The SWE FCs have an variable energy(speed) resolution of $\sim$6.5--13\%($\sim$3.3--6.5\%), depending on energy window.  The ion moments are constrained assuming quasi-neutrality, i.e., $n{\scriptstyle_{e}}$ $=$ $n{\scriptstyle_{p}}$ $+$ 2$n{\scriptstyle_{\alpha}}$, where $n{\scriptstyle_{e}}$ is taken from the WAVES/TNR observations (see explanation below).

\indent  The electron densities ($n{\scriptstyle_{e}}$) and temperatures ($T{\scriptstyle_{e, j}}$, where $j$ $=$ $tot$, $\parallel$, or $\perp$) were taken from the \emph{Wind}/3DP electron electrostatic analyzers \citep[][]{lin95a} with a $\sim$45 or $\sim$78 second cadence and constant energy bin width of $\sim$20\%.  The electron distributions are constructed by merging the data from the EESA Low and EESA High instruments \citep[e.g., see][for instrument details]{wilsoniii09a, wilsoniii10a} from the 3DP suite following a similar method to that in \citet[][]{pulupa14a}.  However, instead of separating the electron distributions into the core, halo, and strahl components, the velocity moments are directly computed from the merged instrument measurements.  The energy range of EESA Low and High are commandable but a notable point is that the lowest energy channel setting is $\sim$3 eV (more commonly the instrument is set to $E{\scriptstyle_{min}}$ $\sim$ 5 eV).  Thus, with a $\sim$20\% energy bin width an approximation of the lowest resolvable temperature would be $\sim$2 eV without fitting to a model function.

\indent  In order to obtain accurate electron moments, EESA measurements must be corrected for the effects of spacecraft floating potential.  We estimate spacecraft potential following the methods in \citet[][]{salem01a}.  We note that unlike many previous missions, \emph{Wind} can unambiguously measure the total electron density by observing the upper hybrid line (also called the plasma line) with the WAVES/TNR instrument \citep[][]{bougeret95a}.  Analysis of the upper hybrid line via the technique of quasi-thermal noise spectroscopy \citep[][]{meyervernet89a} yield an accurate measurement of $n{\scriptstyle_{e}}$ unaffected by spacecraft potential.  The unbiased value of $n{\scriptstyle_{e}}$ measured by TNR is used to validate and refine the spacecraft potential correction, improving accuracy of the electron moments.

\indent  The study is limited to data derived from velocity moments of the entire species over the observed energy ranges, e.g., we do not separate the electron data into the core, halo, or strahl components \citep[e.g.,][]{bale13a, horaites18a, pulupa14a} nor do we account for secondary proton beam contributions \citep[e.g.,][]{wicks16a}.  We also do not separate the distributions by fast or slow solar wind speeds, as this will be addressed in detail in a future study that will also examine the subcomponents of each species [e.g., \textit{Salem et al. in preparation}].

\subsection{Lists and Data Selection}  \label{subsec:ListsandDataSelection}

\indent  The 3DP dataset spans from January 1, 1995 to December 31, 2004, thus we limit all results and analysis to that time range.  As this does not span a solar cycle, we did not perform any solar-cycle-dependent analyses.

\indent  We measure the plasma parameters observed by \emph{Wind} and separate into four categories based upon five constraints (see below for definitions):  all times (\textbf{Constraints 1} and \textbf{2}), all times excluding interplanetary (IP) shocks (\textbf{Constraints 1}--\textbf{3}), only times within magnetic obstacles (MOs) \citep[e.g.,][]{nieveschinchilla16a, nieveschinchilla18a} (\textbf{Constraints 1}, \textbf{2}, and \textbf{4}), and all times excluding MOs (\textbf{Constraints 1}, \textbf{2}, and \textbf{5}).  We define five constraints to limit the ``low quality'' data from the analysis and provide a concise, objective reference for the four categories listed above.  We defined the constraints as:
\begin{description}[itemsep=0pt,parsep=0pt,topsep=0pt]
  \item[\textbf{Constraint 1}]  Require the 3DP and SWE fit flags be $\geq$ 5 and 10, respectively.  The fit flags are statistical estimates of the quality of the model fit results.  The SWE fit flag also indicates whether a constraint was imposed to find a stable solution in the fitting algorithm, i.e., a SWE fit flag $=$ 10 means no assumptions or constraints were necessary.
  \item[\textbf{Constraint 2}]  Require $T{\scriptstyle_{s, tot}}$ $<$ 1 keV \& $T{\scriptstyle_{s, \parallel}}$ $<$ 1.2 keV \& $T{\scriptstyle_{s, \perp}}$ $<$ 1.2 keV \& $\lvert \textbf{\textit{B}}{\scriptstyle_{o,j}} \rvert$ $<$ 120 nT as a second ``high quality'' data qualifier.  This constraint is mostly to remove outliers and extreme conditions that may not be caught by \textbf{Constraint 1}.
  \item[\textbf{Constraint 3}]  Use all time periods excluding IP shocks, where times within/near an IP shock are defined as five hours prior to and one day after the shock arrival at \emph{Wind}.  Time periods defining when the spacecraft was in/around IP shocks are taken from the Harvard Smithsonian Center for Astrophysics' \emph{Wind} shock database\footnote{\url{https://www.cfa.harvard.edu/shocks/wi\_data/}}.
  \item[\textbf{Constraint 4}]  Only use time periods during magnetic obstacles (MOs) within ICMEs.  Time periods defining when the spacecraft was in/around ICMEs are taken from the \emph{Wind} ICME Catalog\footnote{\url{https://wind.nasa.gov/ICMEindex.php}}.  The times within MOs are given by the MO time ranges given for each ICME entry \citep[][]{nieveschinchilla16a, nieveschinchilla18a}.
  \item[\textbf{Constraint 5}]  Use all time periods excluding MOs.
\end{description}
\noindent  Note that all results presented herein satisfy \textbf{Constraints 1} and \textbf{2}.  For the time period analyzed, there were 170 ICMEs and 239 IP shocks.

\indent  Prior to computing any quantity or ratio, we constructed a uniform grid of two minute intervals spanning from January 1, 1995 00:00:33.565 UTC to January 1, 2005 00:00:33.565 UTC.  All data falling within any two minute bin are averaged and from those averages we compute the temperature ratios and plasma betas.  We compute the electron-to-proton, $\left(T{\scriptstyle_{e}}/T{\scriptstyle_{p}}\right){\scriptstyle_{j}}$, electron-to-alpha, $\left(T{\scriptstyle_{e}}/T{\scriptstyle_{\alpha}}\right){\scriptstyle_{j}}$, and alpha-to-proton, $\left(T{\scriptstyle_{\alpha}}/T{\scriptstyle_{p}}\right){\scriptstyle_{j}}$ temperature ratios .  We also compute the electron ($\beta{\scriptstyle_{e, j}}$), proton ($\beta{\scriptstyle_{p, j}}$), and alpha-particle ($\beta{\scriptstyle_{\alpha, j}}$) plasma betas.  Further details and parameter/symbol definitions can be found in Appendix \ref{app:Definitions}.

\indent  The SWE FC instrument does not always fully resolve the proton and alpha-particle peaks.  Thus, the total number of resolved proton and alpha-particle intervals are not the same.  Further, the numerical scheme used to parameterize the distributions generally finds the perpendicular components more easily than the parallel, resulting in more perpendicular than parallel and total component intervals.  Since we do not directly compare perpendicular and parallel or perpendicular and total, we did not eliminate fit results satisfying \textbf{Constraints 1} and \textbf{2} if only the perpendicular solution was valid.

\indent  Note that the electron data set does not include burst data often triggered by solar wind transients (e.g., shocks).  This partly limits the direct comparison between all periods and those excluding interplanetary (IP) shocks for any parameter depending upon electron moments.

\section{Solar Wind Statistics}  \label{sec:SolarWindStatistics}

\indent  In this section we introduce and discuss the one-variable statistics and distributions of $T{\scriptstyle_{s, j}}$, $\left(T{\scriptstyle_{s'}}/T{\scriptstyle_{s}}\right){\scriptstyle_{j}}$, and $\beta{\scriptstyle_{s, j}}$.  In Tables \ref{tab:Temperatures}, \ref{tab:TemperatureRatios}, and \ref{tab:PlasmaBeta} all sections satisfy \textbf{Constraints 1} and \textbf{2}.  The formats of Tables \ref{tab:Temperatures}, \ref{tab:TemperatureRatios}, and \ref{tab:PlasmaBeta} and Figures \ref{fig:Temperatures}, \ref{fig:TemperatureRatios}, and \ref{fig:PlasmaBeta} are all the same, respectively.

\indent  Since none of the parameters have fully Gaussian distributions, we use the median and lower and upper quartile values rather than mean and standard deviation in the tables.  These values more accurately represent the data and are less biased by tails.  The median values are shown in each of the figure panels as well.

\begin{deluxetable*}{| l | c | c | c | c | c | c |}
  \tabletypesize{\small}    
  \tablecaption{Temperature Parameters \label{tab:Temperatures}}
  \tablehead{\colhead{Temperature [eV]} & \colhead{$X{\scriptstyle_{min}}$}\tablenotemark{a} & \colhead{$X{\scriptstyle_{max}}$}\tablenotemark{b} & \colhead{$\bar{\mathbf{X}}$}\tablenotemark{c} & \colhead{$X{\scriptstyle_{25\%}}$}\tablenotemark{d} & \colhead{$X{\scriptstyle_{75\%}}$}\tablenotemark{e} & \colhead{N}\tablenotemark{f}}
  \startdata
  \multicolumn{7}{ |c| }{\textbf{All data in table satisfies Constraints 1 and 2}} \\
  \multicolumn{7}{ |c| }{\textbf{All Good Time Periods}} \\
  \hline
  $T{\scriptstyle_{e, tot}}$ & 2.43 & 58.8 & 11.9 & 10.0 & 14.0 & 820057  \\
  $T{\scriptstyle_{e, \perp}}$ & 2.29 & 77.2 & 12.8 & 10.5 & 15.4 & 820057  \\
  $T{\scriptstyle_{e, \parallel}}$ & 2.49 & 58.6 & 11.4 & 9.7 & 13.4 & 820057  \\
  \hline
  $T{\scriptstyle_{p, tot}}$ & 0.16 & 940.6 & 8.6 & 4.7 & 15.9 & 1095314  \\
  $T{\scriptstyle_{p, \perp}}$ & 0.05 & 1192.7 & 7.5 & 4.2 & 14.7 & 1124159  \\
  $T{\scriptstyle_{p, \parallel}}$ & 0.05 & 1196.1 & 10.7 & 5.2 & 20.0 & 1103884  \\
  \hline
  $T{\scriptstyle_{\alpha, tot}}$ & 0.45 & 964.7 & 10.8 & 4.9 & 31.6 & 476255  \\
  $T{\scriptstyle_{\alpha, \perp}}$ & 0.20 & 1198.6 & 21.4 & 6.5 & 58.0 & 883543  \\
  $T{\scriptstyle_{\alpha, \parallel}}$ & 0.23 & 1192.2 & 14.8 & 5.4 & 50.7 & 564208  \\
  \hline
  \multicolumn{7}{ |c| }{\textbf{Constraint 3:  Time periods excluding IP shocks}} \\
  \hline
  $T{\scriptstyle_{e, tot}}$ & 2.71 & 52.7 & 11.9 & 10.0 & 14.0 & 760815  \\
  $T{\scriptstyle_{e, \perp}}$ & 2.59 & 65.2 & 12.8 & 10.5 & 15.4 & 760815  \\
  $T{\scriptstyle_{e, \parallel}}$ & 2.74 & 46.9 & 11.3 & 9.7 & 13.3 & 760815  \\
  \hline
  $T{\scriptstyle_{p, tot}}$ & 0.16 & 865.8 & 8.6 & 4.7 & 15.8 & 1001957  \\
  $T{\scriptstyle_{p, \perp}}$ & 0.06 & 1192.7 & 7.5 & 4.2 & 14.6 & 1028552  \\
  $T{\scriptstyle_{p, \parallel}}$ & 0.05 & 1173.6 & 10.8 & 5.2 & 19.9 & 1009757  \\
  \hline
  $T{\scriptstyle_{\alpha, tot}}$ & 0.49 & 923.3 & 10.5 & 4.8 & 31.1 & 428536  \\
  $T{\scriptstyle_{\alpha, \perp}}$ & 0.20 & 1198.6 & 21.3 & 6.4 & 57.5 & 804268  \\
  $T{\scriptstyle_{\alpha, \parallel}}$ & 0.23 & 1192.2 & 14.9 & 5.4 & 51.3 & 512190  \\
  \hline
  \multicolumn{7}{ |c| }{\textbf{Constraint 4:  Time periods during ICMEs}} \\
  \hline
  $T{\scriptstyle_{e, tot}}$ & 2.43 & 52.4 & 10.4 & 8.2 & 13.2 & 29389  \\
  $T{\scriptstyle_{e, \perp}}$ & 2.29 & 77.2 & 11.1 & 8.5 & 14.5 & 29389  \\
  $T{\scriptstyle_{e, \parallel}}$ & 2.49 & 52.0 & 10.0 & 7.9 & 12.4 & 29389  \\
  \hline
  $T{\scriptstyle_{p, tot}}$ & 0.39 & 549.5 & 4.2 & 2.5 & 7.5 & 72530  \\
  $T{\scriptstyle_{p, \perp}}$ & 0.06 & 719.6 & 3.9 & 2.3 & 6.9 & 73349  \\
  $T{\scriptstyle_{p, \parallel}}$ & 0.05 & 1196.1 & 4.4 & 2.5 & 8.9 & 73149  \\
  \hline
  $T{\scriptstyle_{\alpha, tot}}$ & 0.49 & 525.1 & 5.8 & 3.1 & 14.5 & 43343  \\
  $T{\scriptstyle_{\alpha, \perp}}$ & 0.25 & 1169.4 & 8.4 & 3.7 & 23.8 & 63344  \\
  $T{\scriptstyle_{\alpha, \parallel}}$ & 0.23 & 1135.5 & 5.8 & 3.0 & 14.2 & 45878  \\
  \hline
  \multicolumn{7}{ |c| }{\textbf{Constraint 5:  Time periods excluding ICMEs}} \\
  \hline
  $T{\scriptstyle_{e, tot}}$ & 2.71 & 58.8 & 11.9 & 10.1 & 14.0 & 790668  \\
  $T{\scriptstyle_{e, \perp}}$ & 2.59 & 65.2 & 12.9 & 10.5 & 15.5 & 790668  \\
  $T{\scriptstyle_{e, \parallel}}$ & 2.74 & 58.6 & 11.4 & 9.7 & 13.4 & 790668  \\
  \hline
  $T{\scriptstyle_{p, tot}}$ & 0.16 & 940.6 & 9.0 & 4.9 & 16.4 & 1022784  \\
  $T{\scriptstyle_{p, \perp}}$ & 0.05 & 1192.7 & 7.9 & 4.4 & 15.2 & 1050810  \\
  $T{\scriptstyle_{p, \parallel}}$ & 0.05 & 1128.5 & 11.3 & 5.6 & 20.5 & 1030735  \\
  \hline
  $T{\scriptstyle_{\alpha, tot}}$ & 0.45 & 964.7 & 11.6 & 5.2 & 33.4 & 432912  \\
  $T{\scriptstyle_{\alpha, \perp}}$ & 0.20 & 1198.6 & 23.2 & 6.9 & 60.3 & 820199  \\
  $T{\scriptstyle_{\alpha, \parallel}}$ & 0.23 & 1192.2 & 16.6 & 5.8 & 53.9 & 518330  \\
  \hline
  \enddata
  \tablenotetext{a}{minimum}
  \tablenotetext{b}{maximum}
  \tablenotetext{c}{median}
  \tablenotetext{d}{lower quartile}
  \tablenotetext{e}{upper quartile}
  \tablenotetext{f}{number of finite measurements}
  \tablecomments{For symbol definitions, see Appendix \ref{app:Definitions}.}
\end{deluxetable*}

\subsection{Plasma Temperatures}  \label{subsec:PlasmaTemperatures}

\indent  In this section we introduce and discuss the one-variable statistics and distributions of $T{\scriptstyle_{s, j}}$ for electrons ($s$ $=$ $e$), protons ($s$ $=$ $p$), and alpha-particles ($s$ $=$ $\alpha$).  Note that the solar wind is a non-equilibrium, weakly collisional, kinetic gas.  Temperatures in such a gas are more representative of the average kinetic energy in the species bulk flow rest frame than the thermodynamic variable.  Thus, we we report temperatures in units of $eV$ rather than $K$.  One can easily convert to kelvin by multiplying any number in $eV$ by $\sim$11,604 $K eV^{-1}$.

\begin{figure*}
  \centering
    {\includegraphics[trim = 0mm 0mm 0mm 0mm, clip, height=170mm]{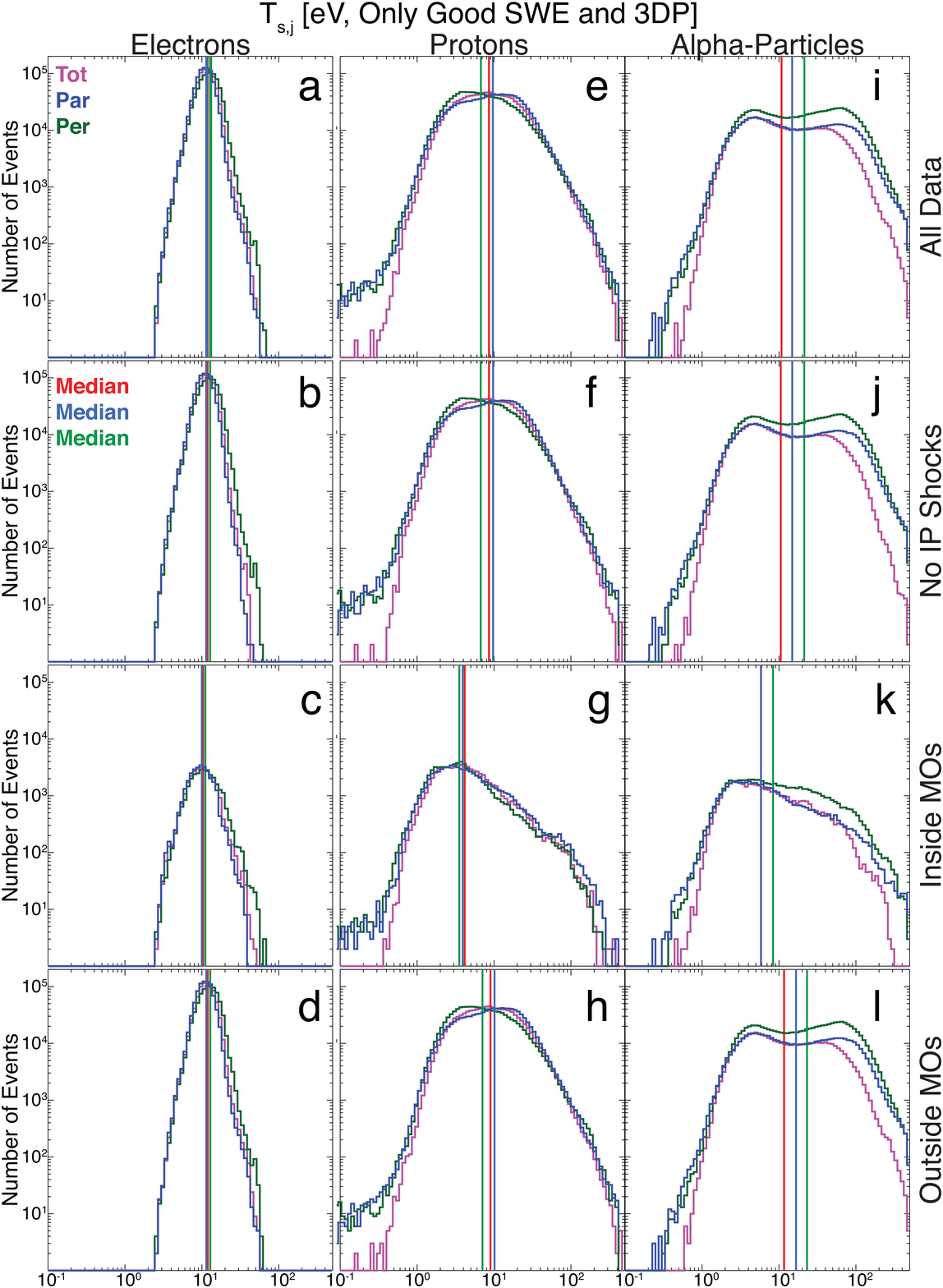}}
    \caption{Temperatures [eV] for different particle species in each column and for the different constraints (i.e., rows) listed in Table \ref{tab:Temperatures}.  In each panel, there are three color-coded histograms for the different components defined as follows: total (magenta); parallel (blue); and perpendicular (green).  The color-coded vertical lines are the median values (values found in Table \ref{tab:Temperatures}) of the distributions for the corresponding color-coded histograms.}
    \label{fig:Temperatures}
\end{figure*}

\indent  Table \ref{tab:Temperatures} and Figure \ref{fig:Temperatures} show the one-variable statistics and histograms of $T{\scriptstyle_{s, j}}$ under the four solar wind categories defined in Section \ref{sec:DefinitionsDataSets}.  The first section of Table \ref{tab:Temperatures} shows the one-variable statistics for all good data.  The ranges for $T{\scriptstyle_{e, j}}$ are much more restricted than for $T{\scriptstyle_{p, j}}$ and $T{\scriptstyle_{\alpha, j}}$.  The mean values (see Supplemental ASCII file) for $T{\scriptstyle_{e, j}}$ and $T{\scriptstyle_{p, j}}$ are similar but the median values $T{\scriptstyle_{e, j}}$ are higher than for $T{\scriptstyle_{p, j}}$.  Interestingly, a comparison between the first two sections show little difference, i.e., the exclusion of IP shocks does not appear to affect the mean or median of any of the temperatures, which is unexpected as ions are often more strongly heated than electrons \citep[][]{wilsoniii10a}.  What does appear to affect the temperatures are the so called magnetic obstacles (MOs) associated with ICMEs \citep[e.g.,][]{nieveschinchilla18a}.  Although this is somewhat expected as the time periods with MOs are partly defined by low $T{\scriptstyle_{p, tot}}$ and/or low $\beta{\scriptstyle_{p, tot}}$, there is rarely a requirement put on the alpha-particle parameters yet they show distinct differences as well.

\indent  Figure \ref{fig:Temperatures} shows color-coded histograms of the total (magenta), parallel (blue), and perpendicular (green) components of $T{\scriptstyle_{s, j}}$ for electrons (first column), protons (second column), and alpha-particles (third column).  Overlaid are color-coded vertical lines showing the median values for each component, under each set of solar wind conditions (separated by rows) listed in Table \ref{tab:Temperatures}.  Note that the rows in Figure \ref{fig:Temperatures} correspond to the sections in Table \ref{tab:Temperatures}.  One can see that $T{\scriptstyle_{p, j}}$ and $T{\scriptstyle_{\alpha, j}}$ have a much broader distribution for all $j$ under all conditions than $T{\scriptstyle_{e, j}}$.  It is not clear why $T{\scriptstyle_{\alpha, j}}$ exhibits a double-peaked distribution at $\sim$4 and $\sim$70 eV for all conditions except during MOs.  We suspect this is related to the interpretation of \citet[][]{maruca13b} where the plasma leaves the solar corona in a super-mass-proportional temperature state and slowly relaxes as it propagates to 1 AU, i.e., the plasma starts with $\left(T{\scriptstyle_{\alpha}}/T{\scriptstyle_{p}}\right){\scriptstyle_{tot}}$ $>$ 4 in the solar corona.

\indent  The majority of temperatures were in the range $\sim$1--50 eV with $\sim$98\% satisfying 5 eV $<$ $T{\scriptstyle_{e, tot}}$ $<$ 20 eV, $\sim$92\% satisfying 1 eV $<$ $T{\scriptstyle_{p, tot}}$ $<$ 30 eV, and $\sim$86\% satisfying 1 eV $<$ $T{\scriptstyle_{p, tot}}$ $<$ 50 eV.  The extrema for $T{\scriptstyle_{s, j}}$ are the minority because the distributions are centrally concentrated.  For instance, only $\sim$0.3\% satisfy $T{\scriptstyle_{e, tot}}$ $<$ 5 eV, in contrast to $\sim$28\% and $\sim$26\%  satisfying $T{\scriptstyle_{p, tot}}$ $<$ 5 eV and $T{\scriptstyle_{\alpha, tot}}$ $<$ 5 eV, respectively.  In the opposite extreme, only $\sim$1.5\% satisfy $T{\scriptstyle_{e, tot}}$ $>$ 20 eV, in contrast to $\sim$2.1\% and $\sim$3.1\%  satisfying $T{\scriptstyle_{p, tot}}$ $>$ 50 eV and $T{\scriptstyle_{\alpha, tot}}$ $>$ 100 eV, respectively.

\subsection{Temperature Ratios}  \label{subsec:TemperatureRatios}

\begin{figure*}
  \centering
    {\includegraphics[trim = 0mm 0mm 0mm 0mm, clip, height=170mm]{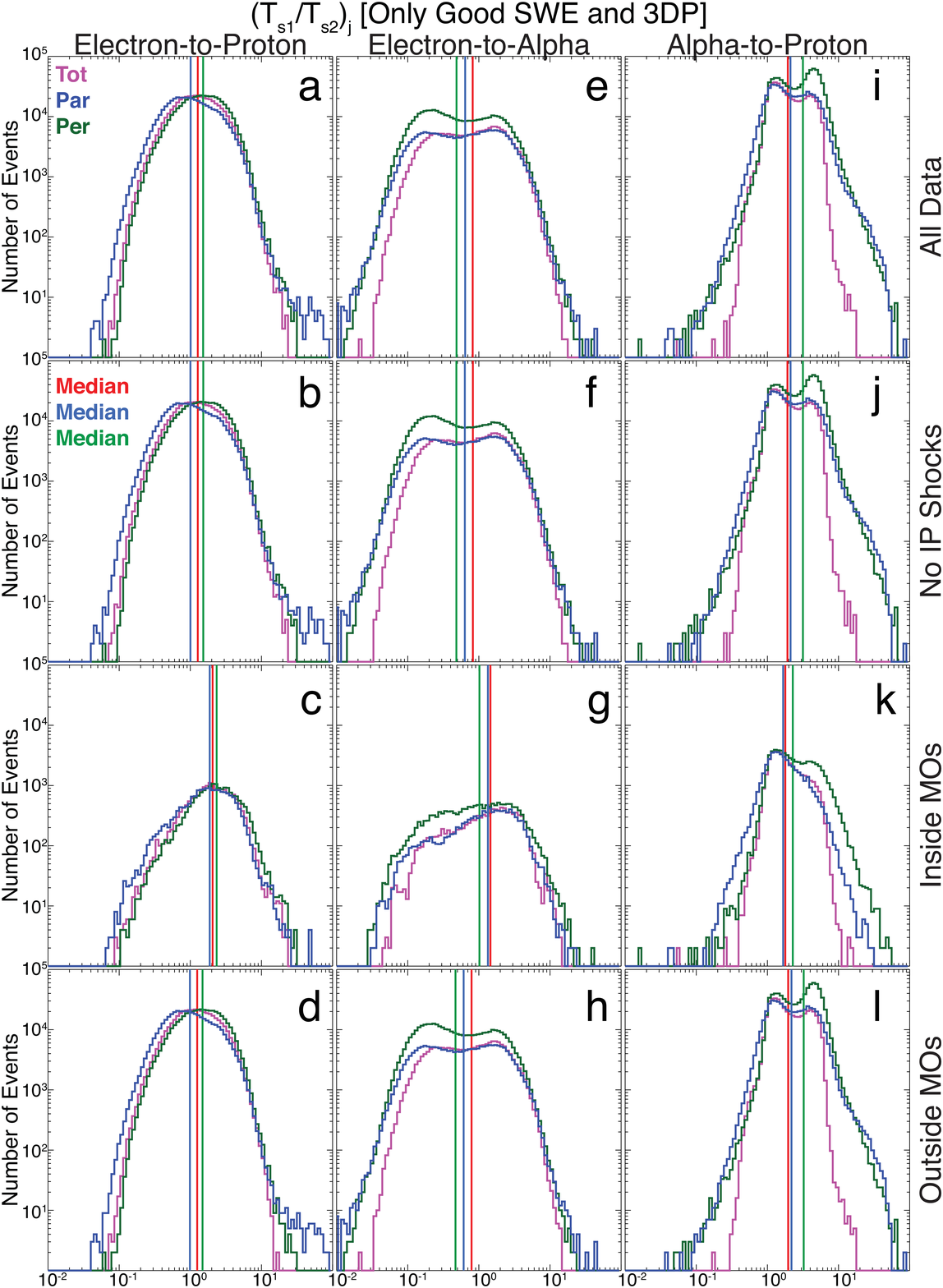}}
    \caption{Temperature ratios [unitless] for different particle species in each column and for the different constraints (i.e., rows) listed in Table \ref{tab:TemperatureRatios}.  In each panel, there are three color-coded histograms for the different components defined as follows: total (magenta); parallel (blue); and perpendicular (green).  The color-coded vertical lines are the median values (values found in Table \ref{tab:TemperatureRatios}) of the distributions for the corresponding color-coded histograms.}
    \label{fig:TemperatureRatios}
\end{figure*}

\indent  In this section we introduce and discuss the one-variable statistics and distributions of $\left(T{\scriptstyle_{s'}}/T{\scriptstyle_{s}}\right){\scriptstyle_{j}}$ for electron-to-proton ($s'$ $=$ $e$ and $s$ $=$ $p$), electron-to-alpha-particle ($s'$ $=$ $e$ and $s$ $=$ $\alpha$), and proton-to-alpha-particle ($s$ $=$ $p$ and $s$ $=$ $\alpha$) ratios.

\begin{deluxetable*}{| l | c | c | c | c | c | c |}
  \tabletypesize{\small}    
  \tablecaption{Temperature Ratio Parameters \label{tab:TemperatureRatios}}
  \tablehead{\colhead{Temperature Ratios [N/A]} & \colhead{$X{\scriptstyle_{min}}$} & \colhead{$X{\scriptstyle_{max}}$} & \colhead{$\bar{\mathbf{X}}$} & \colhead{$X{\scriptstyle_{25\%}}$} & \colhead{$X{\scriptstyle_{75\%}}$} & \colhead{N}}
  \startdata
  \multicolumn{7}{ |c| }{\textbf{All data in table satisfies Constraints 1 and 2}} \\
  \multicolumn{7}{ |c| }{\textbf{All Good Time Periods}} \\
  \hline
  $\left(T{\scriptstyle_{e}}/T{\scriptstyle_{p}}\right){\scriptstyle_{tot}}$ & 0.07 & 25.3 & 1.27 & 0.78 & 2.14 & 445801  \\
  $\left(T{\scriptstyle_{e}}/T{\scriptstyle_{p}}\right){\scriptstyle_{\perp}}$ & 0.03 & 184.6 & 1.51 & 0.90 & 2.49 & 454588  \\
  $\left(T{\scriptstyle_{e}}/T{\scriptstyle_{p}}\right){\scriptstyle_{\parallel}}$ & 0.02 & 161.1 & 1.01 & 0.60 & 1.83 & 445732  \\
  \hline
  $\left(T{\scriptstyle_{e}}/T{\scriptstyle_{\alpha}}\right){\scriptstyle_{tot}}$ & 0.02 & 22.9 & 0.82 & 0.32 & 1.78 & 193704  \\
  $\left(T{\scriptstyle_{e}}/T{\scriptstyle_{\alpha}}\right){\scriptstyle_{\perp}}$ & 0.01 & 42.3 & 0.48 & 0.21 & 1.41 & 393745  \\
  $\left(T{\scriptstyle_{e}}/T{\scriptstyle_{\alpha}}\right){\scriptstyle_{\parallel}}$ & 0.009 & 45.8 & 0.64 & 0.23 & 1.62 & 207313  \\
  \hline
  $\left(T{\scriptstyle_{\alpha}}/T{\scriptstyle_{p}}\right){\scriptstyle_{tot}}$ & 0.02 & 19.0 & 1.94 & 1.35 & 3.49 & 476255  \\
  $\left(T{\scriptstyle_{\alpha}}/T{\scriptstyle_{p}}\right){\scriptstyle_{\perp}}$ & 0.004 & 200.0 & 3.16 & 1.63 & 4.79 & 883427  \\
  $\left(T{\scriptstyle_{\alpha}}/T{\scriptstyle_{p}}\right){\scriptstyle_{\parallel}}$ & 0.02 & 271.7 & 2.11 & 1.33 & 3.87 & 564072  \\
  \hline
  \multicolumn{7}{ |c| }{\textbf{Constraint 3:  Time periods excluding IP shocks}} \\
  \hline
  $\left(T{\scriptstyle_{e}}/T{\scriptstyle_{p}}\right){\scriptstyle_{tot}}$ & 0.07 & 25.3 & 1.27 & 0.78 & 2.14 & 412947  \\
  $\left(T{\scriptstyle_{e}}/T{\scriptstyle_{p}}\right){\scriptstyle_{\perp}}$ & 0.06 & 184.6 & 1.52 & 0.91 & 2.51 & 421080  \\
  $\left(T{\scriptstyle_{e}}/T{\scriptstyle_{p}}\right){\scriptstyle_{\parallel}}$ & 0.02 & 161.1 & 1.00 & 0.60 & 1.82 & 412917  \\
  \hline
  $\left(T{\scriptstyle_{e}}/T{\scriptstyle_{\alpha}}\right){\scriptstyle_{tot}}$ & 0.02 & 22.9 & 0.82 & 0.32 & 1.80 & 177265  \\
  $\left(T{\scriptstyle_{e}}/T{\scriptstyle_{\alpha}}\right){\scriptstyle_{\perp}}$ & 0.01 & 42.3 & 0.48 & 0.21 & 1.42 & 363851  \\
  $\left(T{\scriptstyle_{e}}/T{\scriptstyle_{\alpha}}\right){\scriptstyle_{\parallel}}$ & 0.009 & 45.8 & 0.64 & 0.23 & 1.63 & 190413  \\
  \hline
  $\left(T{\scriptstyle_{\alpha}}/T{\scriptstyle_{p}}\right){\scriptstyle_{tot}}$ & 0.02 & 19.0 & 1.91 & 1.33 & 3.48 & 428536  \\
  $\left(T{\scriptstyle_{\alpha}}/T{\scriptstyle_{p}}\right){\scriptstyle_{\perp}}$ & 0.01 & 200.0 & 3.16 & 1.62 & 4.77 & 804166  \\
  $\left(T{\scriptstyle_{\alpha}}/T{\scriptstyle_{p}}\right){\scriptstyle_{\parallel}}$ & 0.02 & 271.7 & 2.12 & 1.32 & 3.92 & 512077  \\
  \hline
  \multicolumn{7}{ |c| }{\textbf{Constraint 4:  Time periods during ICMEs}} \\
  \hline
  $\left(T{\scriptstyle_{e}}/T{\scriptstyle_{p}}\right){\scriptstyle_{tot}}$ & 0.09 & 25.3 & 2.06 & 1.27 & 3.33 & 18203  \\
  $\left(T{\scriptstyle_{e}}/T{\scriptstyle_{p}}\right){\scriptstyle_{\perp}}$ & 0.1 & 79.4 & 2.35 & 1.46 & 3.76 & 18356  \\
  $\left(T{\scriptstyle_{e}}/T{\scriptstyle_{p}}\right){\scriptstyle_{\parallel}}$ & 0.03 & 161.1 & 1.87 & 1.09 & 3.13 & 18186  \\
  \hline
  $\left(T{\scriptstyle_{e}}/T{\scriptstyle_{\alpha}}\right){\scriptstyle_{tot}}$ & 0.04 & 22.9 & 1.45 & 0.60 & 2.65 & 10077  \\
  $\left(T{\scriptstyle_{e}}/T{\scriptstyle_{\alpha}}\right){\scriptstyle_{\perp}}$ & 0.02 & 42.3 & 1.02 & 0.38 & 2.35 & 16643  \\
  $\left(T{\scriptstyle_{e}}/T{\scriptstyle_{\alpha}}\right){\scriptstyle_{\parallel}}$ & 0.02 & 30.0 & 1.34 & 0.58 & 2.55 & 10290  \\
  \hline
  $\left(T{\scriptstyle_{\alpha}}/T{\scriptstyle_{p}}\right){\scriptstyle_{tot}}$ & 0.02 & 18.3 & 1.78 & 1.34 & 2.80 & 43343  \\
  $\left(T{\scriptstyle_{\alpha}}/T{\scriptstyle_{p}}\right){\scriptstyle_{\perp}}$ & 0.004 & 79.8 & 2.27 & 1.46 & 4.19 & 63318  \\
  $\left(T{\scriptstyle_{\alpha}}/T{\scriptstyle_{p}}\right){\scriptstyle_{\parallel}}$ & 0.04 & 40.3 & 1.66 & 1.22 & 2.58 & 45837  \\
  \hline
  \multicolumn{7}{ |c| }{\textbf{Constraint 5:  Time periods excluding ICMEs}} \\
  \hline
  $\left(T{\scriptstyle_{e}}/T{\scriptstyle_{p}}\right){\scriptstyle_{tot}}$ & 0.07 & 23.2 & 1.25 & 0.77 & 2.09 & 427598  \\
  $\left(T{\scriptstyle_{e}}/T{\scriptstyle_{p}}\right){\scriptstyle_{\perp}}$ & 0.03 & 184.6 & 1.49 & 0.89 & 2.45 & 436232  \\
  $\left(T{\scriptstyle_{e}}/T{\scriptstyle_{p}}\right){\scriptstyle_{\parallel}}$ & 0.02 & 129.3 & 0.99 & 0.59 & 1.77 & 427546  \\
  \hline
  $\left(T{\scriptstyle_{e}}/T{\scriptstyle_{\alpha}}\right){\scriptstyle_{tot}}$ & 0.02 & 17.2 & 0.79 & 0.32 & 1.74 & 183627  \\
  $\left(T{\scriptstyle_{e}}/T{\scriptstyle_{\alpha}}\right){\scriptstyle_{\perp}}$ & 0.01 & 37.6 & 0.47 & 0.21 & 1.37 & 377102  \\
  $\left(T{\scriptstyle_{e}}/T{\scriptstyle_{\alpha}}\right){\scriptstyle_{\parallel}}$ & 0.009 & 45.8 & 0.61 & 0.22 & 1.57 & 197023  \\
  \hline
  $\left(T{\scriptstyle_{\alpha}}/T{\scriptstyle_{p}}\right){\scriptstyle_{tot}}$ & 0.2 & 19.0 & 1.96 & 1.35 & 3.55 & 432912  \\
  $\left(T{\scriptstyle_{\alpha}}/T{\scriptstyle_{p}}\right){\scriptstyle_{\perp}}$ & 0.01 & 200.0 & 3.24 & 1.65 & 4.81 & 820109  \\
  $\left(T{\scriptstyle_{\alpha}}/T{\scriptstyle_{p}}\right){\scriptstyle_{\parallel}}$ & 0.02 & 271.7 & 2.18 & 1.34 & 3.97 & 518235  \\
  \hline
  \enddata
  \tablecomments{Definitions/Symbols are the same as in Table \ref{tab:Temperatures}.  For symbol definitions, see Appendix \ref{app:Definitions}.}
\end{deluxetable*}

\indent  Table \ref{tab:TemperatureRatios} and Figure \ref{fig:TemperatureRatios} show the one-variable statistics of $\left(T{\scriptstyle_{s'}}/T{\scriptstyle_{s}}\right){\scriptstyle_{j}}$ under the four solar wind categories defined in Section \ref{sec:DefinitionsDataSets}.  Similar to $T{\scriptstyle_{s, j}}$, the only solar wind condition that appears to show a significant effect on any of the $\left(T{\scriptstyle_{s'}}/T{\scriptstyle_{s}}\right){\scriptstyle_{j}}$ ratios is the periods during MOs.  The mean and median values for $\left(T{\scriptstyle_{e}}/T{\scriptstyle_{s}}\right){\scriptstyle_{j}}$ are larger during these periods, consistent with MOs showing little effect on $T{\scriptstyle_{e, j}}$ but strongly depressing $T{\scriptstyle_{p, j}}$ and $T{\scriptstyle_{\alpha, j}}$.  The distinctive double-peak in $\left(T{\scriptstyle_{\alpha}}/T{\scriptstyle_{p}}\right){\scriptstyle_{j}}$ in the third column of Figure \ref{fig:TemperatureRatios} is consistent with previous observations \citep[e.g.,][]{kasper06a, kasper08a}.  The first peak near unity would correspond to thermal equilibrium between the two ion species in a collisionally mediated gas.  The second peak near four would correspond to equal thermal speeds, but we suspect this results from super-mass-proportional heating in the corona that has not yet thermalized \citep[][]{maruca13b}.

\indent  In contrast, the electrons almost never exhibit mass-proportional temperatures with either the protons or alpha-particles but $\sim$49\% satisfy 0.5 $<$ $\left(T{\scriptstyle_{e}}/T{\scriptstyle_{p}}\right){\scriptstyle_{tot}}$ $<$ 1.5 while $\sim$32\% of $\left(T{\scriptstyle_{e}}/T{\scriptstyle_{\alpha}}\right){\scriptstyle_{tot}}$ and $\sim$34\% of $\left(T{\scriptstyle_{\alpha}}/T{\scriptstyle_{p}}\right){\scriptstyle_{tot}}$ satisfy the same criteria.  Thus, a sizable fraction of the time the electrons, protons, and alpha-particles are near thermal equilibrium with each other.  The higher ratio for electrons and protons could result from the higher electron-proton than proton-alpha and electron-alpha Coulomb collision rates in the solar wind \citep[e.g.,][]{spitzer53a, borovsky11a} (e.g., see Section \ref{subsec:CollisionRates} and Appendix \ref{app:CoulombCollisions}) or it may be a reflection of the primordial solar wind near the sun \citep[e.g.,][]{kasper17b}.  It is important to note that the collision rate is not the heating or heat-exchange rate.  An interesting aside is that $\left(T{\scriptstyle_{e}}/T{\scriptstyle_{p}}\right){\scriptstyle_{\perp}}$ $>$ $\left(T{\scriptstyle_{e}}/T{\scriptstyle_{p}}\right){\scriptstyle_{\parallel}}$ is satisfied only $\sim$17\% of the time while $\left(T{\scriptstyle_{e}}/T{\scriptstyle_{\alpha}}\right){\scriptstyle_{\perp}}$ $>$ $\left(T{\scriptstyle_{e}}/T{\scriptstyle_{\alpha}}\right){\scriptstyle_{\parallel}}$ is satisfied $\sim$32\% of the time.  However, the median values for $\left(T{\scriptstyle_{s'}}/T{\scriptstyle_{p}}\right){\scriptstyle_{\perp}}$ are larger than $\left(T{\scriptstyle_{s'}}/T{\scriptstyle_{p}}\right){\scriptstyle_{\parallel}}$ ($s'$ $=$ $e$ or $\alpha$) for all four solar wind categories in Table \ref{tab:TemperatureRatios}.  The opposite is true for the electron-to-alpha-particle ratios.  What may not be captured here is the energy-dependent interactions between waves/turbulence and charged particles.  For instance, it is known that electrostatic ion-acoustic waves can generate anisotropic, energetic ion tails while simultaneously generating anisotropic heating of the core electrons \citep[e.g.,][]{dum78a, dum78b}.  The net effects may not significantly alter the total temperatures of either species, as seen in previous studies \citep[e.g.,][]{wilsoniii10a}.  However, a detailed investigation of the heating mechanisms and subcomponents of each species is beyond the scope of this work.

\indent  The double-peak in $\left(T{\scriptstyle_{e}}/T{\scriptstyle_{\alpha}}\right){\scriptstyle_{j}}$ in the second column of Figure \ref{fig:TemperatureRatios} has local maxima near $\sim$0.2 and $\sim$2.0.  This appears to be a consequence of the combination of a double-peaked $T{\scriptstyle_{\alpha, j}}$ spanning the single-peaked $T{\scriptstyle_{e, j}}$ seen in Figure \ref{fig:Temperatures}.

\indent  The majority of data were in the range $\sim$0.5--5.0 with $\sim$88\% for $\left(T{\scriptstyle_{e}}/T{\scriptstyle_{p}}\right){\scriptstyle_{tot}}$, $\sim$61\% for $\left(T{\scriptstyle_{e}}/T{\scriptstyle_{\alpha}}\right){\scriptstyle_{tot}}$, and $\sim$92\% for $\left(T{\scriptstyle_{\alpha}}/T{\scriptstyle_{p}}\right){\scriptstyle_{tot}}$.  Similar to $T{\scriptstyle_{s, j}}$, the extrema for $\left(T{\scriptstyle_{s'}}/T{\scriptstyle_{s}}\right){\scriptstyle_{j}}$ are the minority.  For instance, only $\sim$9.7\% satisfy $\left(T{\scriptstyle_{e}}/T{\scriptstyle_{p}}\right){\scriptstyle_{tot}}$ $<$ 0.5, $\sim$1.5\% satisfy $\left(T{\scriptstyle_{e}}/T{\scriptstyle_{\alpha}}\right){\scriptstyle_{tot}}$ $<$ 0.1, and $\sim$4.4\% satisfy $\left(T{\scriptstyle_{\alpha}}/T{\scriptstyle_{p}}\right){\scriptstyle_{tot}}$ $<$ 1.0.  On the opposite side, $\sim$2.3\%, $\sim$1.7\%, and $\sim$8.0\% satisfy $\left(T{\scriptstyle_{e}}/T{\scriptstyle_{p}}\right){\scriptstyle_{tot}}$ $>$ 5, $\left(T{\scriptstyle_{e}}/T{\scriptstyle_{\alpha}}\right){\scriptstyle_{tot}}$ $>$ 5, and $\left(T{\scriptstyle_{\alpha}}/T{\scriptstyle_{p}}\right){\scriptstyle_{tot}}$ $>$ 5, respectively.

\begin{figure*}
  \centering
    {\includegraphics[trim = 0mm 0mm 0mm 0mm, clip, height=170mm]{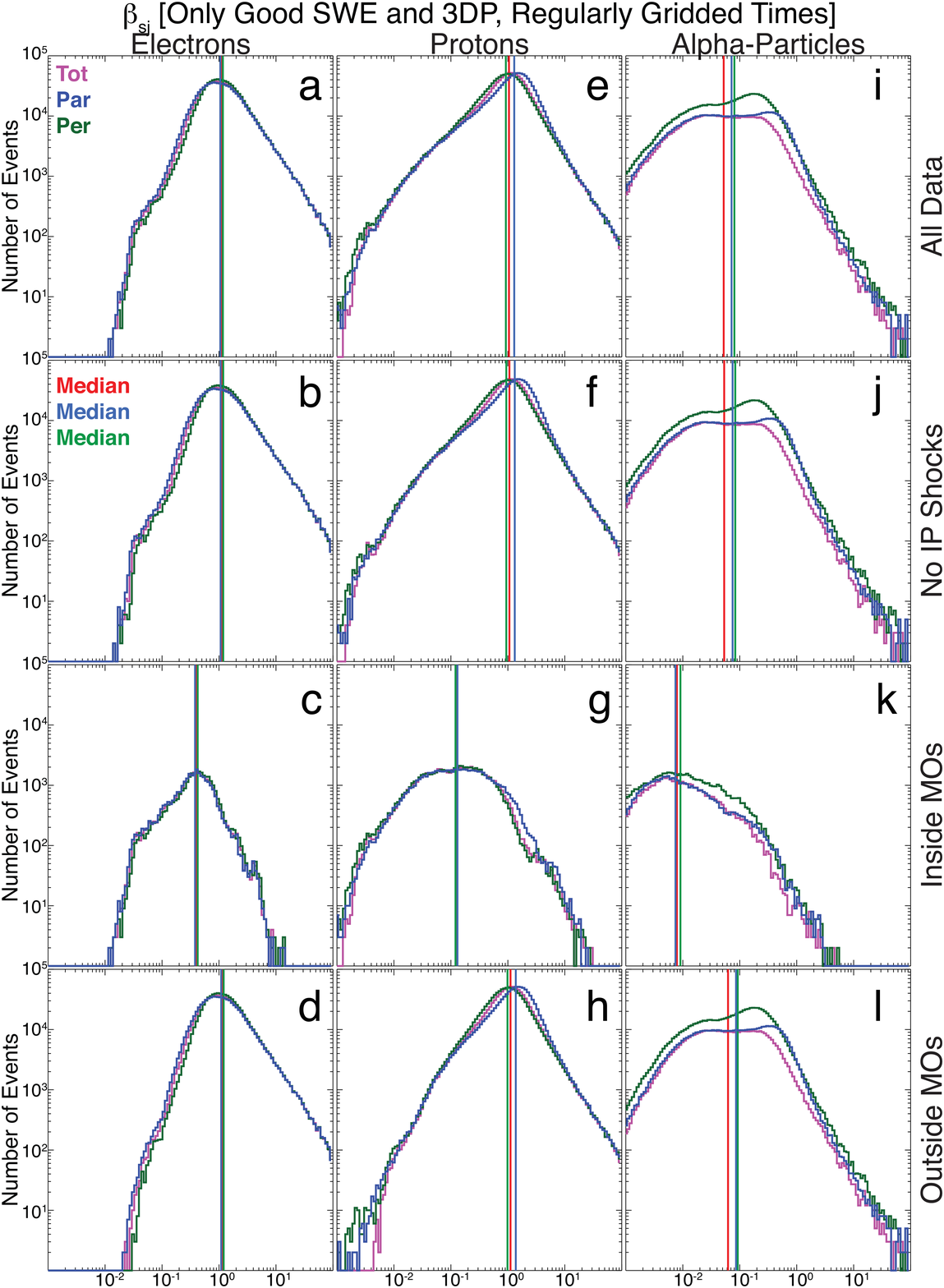}}
    \caption{Plasma betas [unitless] for different particle species in each column and for the different constraints (i.e., rows) listed in Table \ref{tab:PlasmaBeta}.  In each panel, there are three color-coded histograms for the different components defined as follows: total (magenta); parallel (blue); and perpendicular (green).  The color-coded vertical lines are the median values (values found in Table \ref{tab:PlasmaBeta}) of the distributions for the corresponding color-coded histograms.}
    \label{fig:PlasmaBeta}
\end{figure*}

\subsection{Plasma Betas}  \label{subsec:PlasmaBetas}

\indent  In this section we introduce and discuss the one-variable statistics and distributions of $\beta{\scriptstyle_{s, j}}$ for electrons ($s$ $=$ $e$), protons ($s$ $=$ $p$), and alpha-particles ($s$ $=$ $\alpha$).

\indent  Table \ref{tab:PlasmaBeta} and Figure \ref{fig:PlasmaBeta} show the one-variable statistics of $\beta{\scriptstyle_{s, j}}$ under the four solar wind categories defined in Section \ref{sec:DefinitionsDataSets}.  Both $\beta{\scriptstyle_{e, j}}$ and $\beta{\scriptstyle_{p, j}}$ have median values near unity under all conditions except during MOs.  The smaller median values of $\beta{\scriptstyle_{\alpha, j}}$ are likely dominated by the consistently smaller $n{\scriptstyle_{\alpha}}$ values in the solar wind.  Again, all three regions for each $\beta{\scriptstyle_{s, j}}$ look the same except for periods during MOs, which are statistically smaller for all species (by roughly an order of magnitude for protons and alpha-particles).  Curiously, the $\beta{\scriptstyle_{e, j}}$ are much lower during MOs despite the $T{\scriptstyle_{e, j}}$ values being roughly the same under all four solar wind categories implying smaller $n{\scriptstyle_{e}}$ during MOs.

\indent  Note that care should be taken when reading the range of possible $\beta{\scriptstyle_{s, j}}$ under all conditions (i.e., first section in Table \ref{tab:PlasmaBeta}).  Although our stringent criteria outlined in Section \ref{sec:DefinitionsDataSets} were intended to remove unphysical results, such extremes in $\beta{\scriptstyle_{s, j}}$ should be viewed with additional scrutiny.  For instance, only $\sim$0.5\% satisfy $\beta{\scriptstyle_{e, tot}}$ $<$ 0.1, $\sim$4.9\% satisfy $\beta{\scriptstyle_{p, tot}}$ $<$ 0.1, and $\sim$0.6\% satisfy $\beta{\scriptstyle_{\alpha, tot}}$ $<$ 0.001.  In contrast, only $\sim$2.6\% satisfy $\beta{\scriptstyle_{e, tot}}$ $>$ 10, $\sim$1.5\% satisfy $\beta{\scriptstyle_{p, tot}}$ $>$ 10, and $\sim$1.8\% satisfy $\beta{\scriptstyle_{\alpha, tot}}$ $>$ 1.  The majority of electron and proton betas were in the range $\sim$0.5--5.0 with $\sim$78\% satisfying 0.5 $<$ $\beta{\scriptstyle_{e, tot}}$ $<$ 5.0, $\sim$73\% satisfying 0.5 $<$ $\beta{\scriptstyle_{p, tot}}$ $<$ 5.0, while $\sim$78\% of the alpha-particle betas satisfied 0.01 $<$ $\beta{\scriptstyle_{\alpha, tot}}$ $<$ 0.5.

\begin{deluxetable*}{| l | c | c | c | c | c | c |}
  \tabletypesize{\small}    
  \tablecaption{Plasma Beta Parameters \label{tab:PlasmaBeta}}
  \tablehead{\colhead{Plasma Betas [N/A]} & \colhead{$X{\scriptstyle_{min}}$} & \colhead{$X{\scriptstyle_{max}}$} & \colhead{$\bar{\mathbf{X}}$} & \colhead{$X{\scriptstyle_{25\%}}$} & \colhead{$X{\scriptstyle_{75\%}}$} & \colhead{N}}
  \startdata
  \multicolumn{7}{ |c| }{\textbf{All data in table satisfies Constraints 1 and 2}} \\
  \multicolumn{7}{ |c| }{\textbf{All Good Time Periods}} \\
  \hline
  $\beta{\scriptstyle_{e, tot}}$ & 0.006 & 8870 & 1.09 & 0.64 & 1.99 & 820056  \\
  $\beta{\scriptstyle_{e, \perp}}$ & 0.007 & 8914 & 1.17 & 0.71 & 2.06 & 820056  \\
  $\beta{\scriptstyle_{e, \parallel}}$ & 0.005 & 8848 & 1.05 & 0.60 & 1.96 & 820056  \\
  \hline
  $\beta{\scriptstyle_{p, tot}}$ & 0.001 & 4568 & 1.05 & 0.54 & 1.77 & 1095171  \\
  $\beta{\scriptstyle_{p, \perp}}$ & $4 \times 10^{-5}$ & 4391 & 0.92 & 0.47 & 1.62 & 1124001  \\
  $\beta{\scriptstyle_{p, \parallel}}$ & $6 \times 10^{-5}$ & 4923 & 1.29 & 0.62 & 2.18 & 1103741  \\
  \hline
  $\beta{\scriptstyle_{\alpha, tot}}$ & $5 \times 10^{-5}$ & 612 & 0.05 & 0.02 & 0.17 & 476129  \\
  $\beta{\scriptstyle_{\alpha, \perp}}$ & $5 \times 10^{-5}$ & 594 & 0.08 & 0.02 & 0.22 & 883409  \\
  $\beta{\scriptstyle_{\alpha, \parallel}}$ & $2 \times 10^{-5}$ & 647 & 0.07 & 0.02 & 0.27 & 564081  \\
  \hline
  \multicolumn{7}{ |c| }{\textbf{Constraint 3:  Time periods excluding IP shocks}} \\
  \hline
  $\beta{\scriptstyle_{e, tot}}$ & 0.006 & 8870 & 1.11 & 0.65 & 2.01 & 760814  \\
  $\beta{\scriptstyle_{e, \perp}}$ & 0.007 & 8914 & 1.19 & 0.73 & 2.07 & 760814  \\
  $\beta{\scriptstyle_{e, \parallel}}$ & 0.005 & 8848 & 1.07 & 0.61 & 1.98 & 760814  \\
  \hline
  $\beta{\scriptstyle_{p, tot}}$ & 0.001 & 4568 & 1.07 & 0.57 & 1.78 & 1001816  \\
  $\beta{\scriptstyle_{p, \perp}}$ & $8 \times 10^{-5}$ & 4391 & 0.94 & 0.49 & 1.63 & 1028396  \\
  $\beta{\scriptstyle_{p, \parallel}}$ & 0.0001 & 4923 & 1.32 & 0.66 & 2.20 & 1009616  \\
  \hline
  $\beta{\scriptstyle_{\alpha, tot}}$ & $5 \times 10^{-5}$ & 295 & 0.05 & 0.02 & 0.18 & 428411  \\
  $\beta{\scriptstyle_{\alpha, \perp}}$ & $8 \times 10^{-5}$ & 373 & 0.08 & 0.02 & 0.22 & 804136  \\
  $\beta{\scriptstyle_{\alpha, \parallel}}$ & $2 \times 10^{-5}$ & 415 & 0.07 & 0.02 & 0.28 & 512064  \\
  \hline
  \multicolumn{7}{ |c| }{\textbf{Constraint 4:  Time periods during ICMEs}} \\
  \hline
  $\beta{\scriptstyle_{e, tot}}$ & 0.006 & 37 & 0.40 & 0.23 & 0.62 & 29389  \\
  $\beta{\scriptstyle_{e, \perp}}$ & 0.007 & 38 & 0.42 & 0.25 & 0.66 & 29389  \\
  $\beta{\scriptstyle_{e, \parallel}}$ & 0.005 & 36 & 0.38 & 0.22 & 0.60 & 29389  \\
  \hline
  $\beta{\scriptstyle_{p, tot}}$ & 0.001 & 159 & 0.13 & 0.05 & 0.31 & 72530  \\
  $\beta{\scriptstyle_{p, \perp}}$ & $4 \times 10^{-5}$ & 157 & 0.12 & 0.04 & 0.29 & 73349  \\
  $\beta{\scriptstyle_{p, \parallel}}$ & $6 \times 10^{-5}$ & 164 & 0.13 & 0.05 & 0.35 & 73149  \\
  \hline
  $\beta{\scriptstyle_{\alpha, tot}}$ & $5 \times 10^{-5}$ & 29 & 0.01 & 0.00 & 0.02 & 43343  \\
  $\beta{\scriptstyle_{\alpha, \perp}}$ & $5 \times 10^{-5}$ & 49 & 0.01 & 0.00 & 0.03 & 63344  \\
  $\beta{\scriptstyle_{\alpha, \parallel}}$ & $2 \times 10^{-5}$ & 32 & 0.01 & 0.00 & 0.02 & 45878  \\
  \hline
  \multicolumn{7}{ |c| }{\textbf{Constraint 5:  Time periods excluding ICMEs}} \\
  \hline
  $\beta{\scriptstyle_{e, tot}}$ & 0.02 & 8870 & 1.13 & 0.67 & 2.04 & 790667  \\
  $\beta{\scriptstyle_{e, \perp}}$ & 0.03 & 8914 & 1.21 & 0.74 & 2.10 & 790667  \\
  $\beta{\scriptstyle_{e, \parallel}}$ & 0.02 & 8848 & 1.09 & 0.63 & 2.01 & 790667  \\
  \hline
  $\beta{\scriptstyle_{p, tot}}$ & 0.002 & 4568 & 1.11 & 0.63 & 1.85 & 1022641  \\
  $\beta{\scriptstyle_{p, \perp}}$ & 0.0003 & 4391 & 0.98 & 0.54 & 1.69 & 1050652  \\
  $\beta{\scriptstyle_{p, \parallel}}$ & 0.0001 & 4923 & 1.38 & 0.73 & 2.27 & 1030592  \\
  \hline
  $\beta{\scriptstyle_{\alpha, tot}}$ & 0.0002 & 612 & 0.06 & 0.02 & 0.19 & 432786  \\
  $\beta{\scriptstyle_{\alpha, \perp}}$ & $5 \times 10^{-5}$ & 594 & 0.09 & 0.03 & 0.23 & 820065  \\
  $\beta{\scriptstyle_{\alpha, \parallel}}$ & $6 \times 10^{-5}$ & 647 & 0.09 & 0.02 & 0.29 & 518203  \\
  \hline
  \enddata
  \tablecomments{Definitions/Symbols are the same as in Table \ref{tab:Temperatures}.  For symbol definitions, see Appendix \ref{app:Definitions}.}
\end{deluxetable*}

\subsection{Collision Rates}  \label{subsec:CollisionRates}

\indent  In this section we introduce and discuss the one-variable statistics and distributions of the classical binary Coulomb collision frequency \citep[e.g.,][]{krall73a, schunk75a, schunk77a, spitzer53a} for a 90$^{\circ}$ deflection angle between particles of species $s$ and $s'$ (see Appendix \ref{app:CoulombCollisions} for definitions and details).  We also estimate the effective wave-particle interaction rates between electrostatic ion-acoustic waves and particles \citep[e.g., see][and references therein]{wilsoniii14a, wilsoniii14b}.  We use electrostatic ion-acoustic waves, instead of other modes, for four reasons: they are common in the solar wind \citep[e.g.,][]{gurnett79b}, they interact with both ions and electrons \citep[e.g.,][]{dum78a, dum78b}, to compare to previous work \citep[e.g.,][]{wilsoniii07a, wilsoniii14b}, and because there is an analytical expression for the collision rate comprised of measurable parameters.  The purpose is to compare with Coulomb collision rates to determine whether wave-particle interactions should be considered in modeling the evolution of the solar wind from the sun to the Earth.

\indent  When we calculated the particle-particle and wave-particle collision rates for all data satisfying \textbf{Constraints 1} and \textbf{2} finding the minimum to maximum ranges:

\begin{itemize}[itemsep=0pt,parsep=0pt,topsep=0pt]
  \item[]  \textbf{Range of Values}
  \begin{itemize}[itemsep=0pt,parsep=0pt,topsep=0pt]
    \item $\nu{\scriptstyle_{ee}}$ $\sim$ $1 \times 10^{-8}$ -- $1 \times 10^{-4}$ \# $s^{-1}$
    \item $\nu{\scriptstyle_{pp}}$ $\sim$ $6 \times 10^{-12}$ -- $3 \times 10^{-5}$ \# $s^{-1}$
    \item $\nu{\scriptstyle_{\alpha \alpha}}$ $\sim$ $1 \times 10^{-11}$ -- $8 \times 10^{-6}$ \# $s^{-1}$
    \item $\nu{\scriptstyle_{ep}}$ $\sim$ $4 \times 10^{-8}$ -- $1 \times 10^{-4}$ \# $s^{-1}$
    \item $\nu{\scriptstyle_{e \alpha}}$ $\sim$ $2 \times 10^{-8}$ -- $5 \times 10^{-5}$ \# $s^{-1}$
    \item $\nu{\scriptstyle_{p \alpha}}$ $\sim$ $3 \times 10^{-11}$ -- $3 \times 10^{-6}$ \# $s^{-1}$
  \end{itemize}
  \item[]  \textbf{Median Values}
  \begin{itemize}[itemsep=0pt,parsep=0pt,topsep=0pt]
    \item $\nu{\scriptstyle_{ee}}$ $\sim$ $2 \times 10^{-6}$ \# $s^{-1}$
    \item $\nu{\scriptstyle_{pp}}$ $\sim$ $1 \times 10^{-7}$ \# $s^{-1}$
    \item $\nu{\scriptstyle_{\alpha \alpha}}$ $\sim$ $2 \times 10^{-8}$ \# $s^{-1}$
    \item $\nu{\scriptstyle_{ep}}$ $\sim$ $3 \times 10^{-6}$ \# $s^{-1}$
    \item $\nu{\scriptstyle_{e \alpha}}$ $\sim$ $4 \times 10^{-7}$ \# $s^{-1}$
    \item $\nu{\scriptstyle_{p \alpha}}$ $\sim$ $1 \times 10^{-8}$ \# $s^{-1}$
  \end{itemize}
\end{itemize}

\indent  If we compare the median values for different time periods, we find that the maximum collision rates for all rates defined by Equations \ref{eq:SHcollfreq_0}--\ref{eq:SHcollfreq_5} occur during MOs (i.e., \textbf{Constraint 4}).  See the Supplemental ASCII files for the full statistical results.

\indent  If we compare these rates to the quasi-linear approximation for the effective collision rates \citep[e.g., see Equation \ref{eq:WPIcollfreq_0} and][and references therein]{wilsoniii14a, wilsoniii14b} between ion-acoustic waves and particles, $\nu{\scriptstyle_{iaw}}$, we find $\nu{\scriptstyle_{iaw}}$ $\sim$ $6 \times 10^{-5}$ -- $3 \times 10^{-3}$ \# $s^{-1}$ with a median of $\sim 5 \times 10^{-4}$ \# $s^{-1}$.  Here we used the typical amplitudes observed in the solar wind of $\sim$0.1 mV/m \citep[e.g.,][]{gurnett79b}.  Note for every order of magnitude change in the electric field amplitude, $\nu{\scriptstyle_{iaw}}$ will change by two orders of magnitude.  We should also note that Equation \ref{eq:WPIcollfreq_0} is known to give rates that $\sim$2--3 orders of magnitude too small \citep[e.g.,][]{petkaki08a, yoon06a}.  We still use this equation, however, because it always underestimates the net effect of the waves so it can serve as a lower bound for wave-particle interaction rates.  If we look at the ratio of the ion-acoustic to Coulomb collision rates, we find:

\begin{itemize}[itemsep=0pt,parsep=0pt,topsep=0pt]
  \item[]  \textbf{Range of Values (assuming 0.1 mV/m)}
  \begin{itemize}[itemsep=0pt,parsep=0pt,topsep=0pt]
    \item  $\nu{\scriptstyle_{iaw}}/\nu{\scriptstyle_{ee}}$ $\sim$ $3 \times 10^{+0}$ --  $\sim$ $1 \times 10^{+5}$
    \item  $\nu{\scriptstyle_{iaw}}/\nu{\scriptstyle_{pp}}$ $\sim$ $3 \times 10^{+1}$ --  $\sim$ $9 \times 10^{+6}$
    \item  $\nu{\scriptstyle_{iaw}}/\nu{\scriptstyle_{\alpha \alpha}}$ $\sim$ $1 \times 10^{+2}$ --  $\sim$ $3 \times 10^{+7}$
    \item  $\nu{\scriptstyle_{iaw}}/\nu{\scriptstyle_{ep}}$ $\sim$ $4 \times 10^{+0}$ --  $\sim$ $2 \times 10^{+4}$
    \item  $\nu{\scriptstyle_{iaw}}/\nu{\scriptstyle_{e \alpha}}$ $\sim$ $1 \times 10^{+1}$ --  $\sim$ $2 \times 10^{+4}$
    \item  $\nu{\scriptstyle_{iaw}}/\nu{\scriptstyle_{p \alpha}}$ $\sim$ $2 \times 10^{+2}$ --  $\sim$ $1 \times 10^{+7}$
  \end{itemize}
  \item[]  \textbf{Median Values (assuming 0.1 mV/m)}
  \begin{itemize}[itemsep=0pt,parsep=0pt,topsep=0pt]
    \item  $\nu{\scriptstyle_{iaw}}/\nu{\scriptstyle_{ee}}$ $\sim$ $2 \times 10^{+2}$
    \item  $\nu{\scriptstyle_{iaw}}/\nu{\scriptstyle_{pp}}$ $\sim$ $4 \times 10^{+3}$
    \item  $\nu{\scriptstyle_{iaw}}/\nu{\scriptstyle_{\alpha \alpha}}$ $\sim$ $3 \times 10^{+4}$
    \item  $\nu{\scriptstyle_{iaw}}/\nu{\scriptstyle_{ep}}$ $\sim$ $2 \times 10^{+2}$
    \item  $\nu{\scriptstyle_{iaw}}/\nu{\scriptstyle_{e \alpha}}$ $\sim$ $1 \times 10^{+3}$
    \item  $\nu{\scriptstyle_{iaw}}/\nu{\scriptstyle_{p \alpha}}$ $\sim$ $4 \times 10^{+4}$
  \end{itemize}
\end{itemize}

\noindent  Thus, the wave-particle collision rates are $\sim$3--$3 \times 10^{+7}$ times larger than the Coulomb collision rates, even when considering very low amplitudes.  We will discuss this further in Section \ref{sec:DiscussionandConclusions}.

\section{Discussion and Conclusions}  \label{sec:DiscussionandConclusions}

\indent  We present a long-duration (i.e., over a span of $\sim$10 years) statistical analysis of the temperatures, temperature ratios, and plasma betas of electrons, protons, and alpha-particles observed by the \emph{Wind} spacecraft near 1 AU.  The primary purpose of this study is to provide a convenient and comprehensive statistical summary of the electron, proton, and alpha-particle plasma parameters\footnote{It is not the purpose of this work to detail the properties of the subcomponents of each particle species (e.g., halo electrons, proton beams, etc.), which is beyond the scope of this study and some of which will be addressed in detail in a future study [e.g., \textit{Salem et al. in preparation}].  We also did not provide an in-depth, physical interpretation of the plasma parameters as that will also be addressed in future work.} near 1 AU.  The last long-duration (i.e., more than 1 year) statistical study comparing electrons with protons was published 20 years ago \citep[i.e.,][]{newbury98a} and relied upon only 18 months of $\sim$5 minute averaged data (i.e., $\sim$160,000 measurements).  This study uses nearly 10 years of data spanning from the end of solar cycle 22 through much of solar cycle 23 (i.e., $>$1,000,000 measurements) with time periods separated into four categories (see Section \ref{sec:DefinitionsDataSets} for definitions):  all times (\textbf{Constraints 1} and \textbf{2}), all times excluding interplanetary (IP) shocks (\textbf{Constraints 1}--\textbf{3}), only times within magnetic obstacles (MOs) \citep[e.g.,][]{nieveschinchilla16a, nieveschinchilla18a} (\textbf{Constraints 1}, \textbf{2}, and \textbf{4}), and all times excluding MOs (\textbf{Constraints 1}, \textbf{2}, and \textbf{5}).  Below we discuss the observations and present the conclusions.

\subsection{Discussion}  \label{subsec:Discussion}

\indent  Tables \ref{tab:Temperatures}, \ref{tab:TemperatureRatios}, and \ref{tab:PlasmaBeta} and Figures \ref{fig:Temperatures}, \ref{fig:TemperatureRatios}, and \ref{fig:PlasmaBeta} show that the only time periods that appear to significantly affect the parameters are those during MOs, not IP shocks.  In fact, the only parameter that does not appear to show a significant difference inside vs. outside of MOs are the $T{\scriptstyle_{e, j}}$ data.  This causes the electron-dependent temperature ratios to be higher on average during MOs than the ion-only ratios, since the both the proton and alpha-particle temperatures decrease during MOs.  That the electron temperatures are not significantly affected by MOs may be indicative of their higher mobility/conductivity than the ions.  However, their betas are lower so either the densities drop or magnetic fields increase (most likely the latter based on the MO definition criteria).

\indent  The median scalar temperatures (see Table \ref{tab:Temperatures}) and plasma betas (see Table \ref{tab:PlasmaBeta}) are consistent with previous studies (e.g., see Table \ref{tab:PrevTemperatures} and \ref{tab:PrevBetas} for references) but the median temperature ratios (see Table \ref{tab:TemperatureRatios}) are lower than most previous observations (see Table \ref{tab:PrevTempRatios} for references), despite sharing similar observed data ranges for each ratio.  The differences are magnified when we compare the temperature ratios during MOs (i.e., \textbf{Constraint 4}) to previous studies that focused on ICMEs and MOs.  For instance, previous studies reported much larger average values for $\left(T{\scriptstyle_{e}}/T{\scriptstyle_{p}}\right){\scriptstyle_{tot}}$ (i.e., $\sim$4--6) than these observations (i.e., $\sim$1.6$\pm$1.3).  The electron velocity moments from previous studies may have been calculated over reduced energy ranges, which could account for some of this difference.  The spacecraft potential, which must be corrected for in order to obtain accurate electron moments, may also have have been less precisely measured in prior missions.

\indent  Many previous studies used instruments that had energy ranges consistent with only EESA Low (i.e., $\lesssim$ 1 keV) while this work includes EESA High (i.e., $\lesssim$ 30 keV).  Further, only a few missions, including \emph{Wind}, have been able to accurately measure the upper hybrid line providing an unambiguous value for the total electron density, $n{\scriptstyle_{e}}$.  The values of $n{\scriptstyle_{s}}$ (see Section \ref{sec:DefinitionsDataSets} for details) are used to compute $T{\scriptstyle_{s, j}}$ from elements of the diagonalized pressure tensor of species $s$, assuming an ideal gas law, in nearly all velocity moment software, both past and present (e.g., larger $n{\scriptstyle_{s}}$ for the same mean kinetic energy density will result in smaller $T{\scriptstyle_{s, j}}$).  Thus, more accurate $n{\scriptstyle_{s}}$ should produce more accurate $T{\scriptstyle_{s, j}}$.  We suspect that the improved accuracy of the $n{\scriptstyle_{e}}$, and thus $n{\scriptstyle_{p}}$ and $n{\scriptstyle_{\alpha}}$, improved the temperatures values over some previous work which has altered the corresponding temperature ratios.  For example, this can be seen when one examines the study by \citet[][]{skoug00a}, where their average values for $T{\scriptstyle_{e, tot}}$, $T{\scriptstyle_{p, tot}}$, and $\left(T{\scriptstyle_{e}}/T{\scriptstyle_{p}}\right){\scriptstyle_{tot}}$ inside ICMEs were $\sim$9.2 eV, $\sim$2.0 eV, and $\sim$6.8, respectively, compared to these average values of $\sim$11.1 eV, $\sim$7.7 eV, and $\sim$2.6, respectively.

\indent  As we showed, the wave-particle collision rates are $\sim$$10^{+0}$--$10^{+7}$ times larger than the Coulomb collision rates, even when we use very low amplitudes.  It is known that ion-acoustic wave amplitudes can be more than three orders of magnitude larger than the 0.1 mV/m used here \citep[e.g.,][]{wilsoniii07a, wilsoniii10a, wilsoniii14b}, which would increase $\nu{\scriptstyle_{iaw}}$ by six orders of magnitude.  A potential application of these statistical observations is for instability analysis.  For instance, the growth rate threshold was predicted to critically depend upon the $\left(T{\scriptstyle_{e}}/T{\scriptstyle_{p}}\right){\scriptstyle_{tot}}$ ratio for electrostatic ion-acoustic waves\footnote{Note, however, that temperature gradients, heat fluxes, and other non-Maxwellian velocity distribution features can reduce this threshold significantly \citep[e.g.,][]{dum78a, dum78b}.}, specifically growth is only supposed to occur when $\left(T{\scriptstyle_{e}}/T{\scriptstyle_{p}}\right){\scriptstyle_{tot}}$ $\gtrsim$ 3.  We find this criterion is satisfied $\sim$12.4\% of the time\footnote{We suspect that the fraction of time satisfying $\left(T{\scriptstyle_{e}}/T{\scriptstyle_{p}}\right){\scriptstyle_{tot}}$ $\gtrsim$ 3 would change with an increased time resolution \citep[e.g.,][]{paschmann98a}.} for data satisfying \textbf{Constraints 1} and \textbf{2}.  This is important since electrostatic ion-acoustic waves are known to interact with both the electrons and ions and are observed ubiquitously in the solar wind \citep[e.g.,][]{gurnett79b}, at current sheets \citep[e.g.,][]{malaspina13a}, and at IP shocks \citep[e.g.,][]{wilsoniii07a}.

\indent  The collision rates from Appendix \ref{app:CoulombCollisions} show that even for conservatively low wave amplitudes, the median values of the ratio between the wave-particle and particle-particle collision rates range from $\sim$$10^{+0}$--$10^{+7}$.  If we assume ion-acoustic waves occur whenever $\left(T{\scriptstyle_{e}}/T{\scriptstyle_{p}}\right){\scriptstyle_{tot}}$ $\gtrsim$ 3 (i.e., $\sim$12.4\% occurrence rate), then we can adjust the collision rate ratios by this fractional occurrence rate to find the ratios of the net effects from either type of collision\footnote{Note that the range of collision rate ratios $\sim$$10^{+0}$--$10^{+7}$ assumes ion-acoustic waves exist 100\% of the time a particle distribution streams from the sun to the Earth.  Adjusting by this $\sim$12.4\% is an approximation used to reduce their net effect to more realistic values.}.  Then the median ratios would change to $\sim$$10^{-1}$--$10^{+6}$.  To ensure that $\nu{\scriptstyle_{iaw}}$ is always greater than $\nu{\scriptstyle_{ss'}}$ with the $\sim$12.4\% occurrence rate correction, the wave amplitudes would need only increase to $\sim$0.17 mV/m, consistent with numerous solar wind observations \citep[e.g.,][]{gurnett79b, malaspina13a}.

\indent  Suppose we further limit the occurrence rate by the rate of current sheet crossings near 1 AU.  \citet[][]{malaspina13a} reported that roughly 942 current sheets were observed per day, which is a $\sim$1.1\% occurrence rate.  If we couple that with the $\sim$12.4\% rate for satisfying $\left(T{\scriptstyle_{e}}/T{\scriptstyle_{p}}\right){\scriptstyle_{tot}}$ $\gtrsim$ 3, then we have a $\sim$0.14\% net occurrence rate.  This would reduce the collision rate ratios from $\sim$$10^{+0}$--$10^{+7}$ down to $\sim$$10^{-3}$--$10^{+4}$.  Using the same logic as before, these ratios would approach unity if the wave amplitudes were increased to at least $\sim$1.7 mV/m.  Again, this is not an unrealistically large amplitude for ion-acoustic waves in the solar wind \citep[e.g.,][]{gurnett79b, malaspina13a} and much smaller than those observed at IP shocks \citep[e.g.,][]{wilsoniii07a, wilsoniii10a}.

\indent  The primary limitation of assuming a collision rate between particles and ion-acoustic waves is the lack of a statistical study of the true occurrence rate and amplitudes of such modes in the solar wind.  The closest study was performed nearly 40 years ago by \citet[][]{gurnett79b} using dynamic spectra measurements of solar wind electric and magnetic fields from \emph{Helios} and \emph{Voyager}.  It is well known that dynamic spectra measurements underestimate wave amplitudes by up at least an order of magnitude when the waves are composed of short duration, bursty wave packets \citep[e.g.,][]{tsurutani09a}.  Thus, the ratios above would increase by two orders of magnitude for instantaneous wave amplitudes.  Further, the wave-particle collision rates from Equation \ref{eq:WPIcollfreq_0} are known to be $\sim$2--3 orders of magnitude too small \citep[e.g.,][]{petkaki08a, yoon06a}.  If we only include the collision rate correction, then the $\sim$1.7 mV/m amplitude would reduce to $\sim$0.017 mV/m to match the net effects contributed by Coulomb collisions, even with the $\sim$0.14\% net occurrence rate correction factor.

\indent  Perhaps another way to express the differences is to revisit the collision rate ratios.  If we include the $\sim$0.14\% net occurrence rate correction factor but increase the contributions from wave-particle collisions by $\sim$3 orders of magnitude (i.e., to account for known underestimations in theory and observations), the ratios increase to $\sim$4--$3 \times 10^{+7}$.  That is, there would need to be at least three Coulomb collisions for every collision with an ion-acoustic wave, which would correspond to roughly 11 days\footnote{Estimate takes the inverse of the median Coulomb collision rates from the list as a time scale for one collision, multiplies by number of necessary collisions, then divides by 86,400 seconds/day.} for the fastest Coulomb collision rate and $\sim$2380 days for the slowest.

\subsection{Conclusion}  \label{subsec:Conclusion}

\indent  We summarize the observations by showing the mean(median) [standard deviation] for each parameter for all time periods (i.e., satisfying \textbf{Constraints 1} and \textbf{2}) analyzed.  Note that none of the distributions are symmetric (i.e., they all have a finite skewness), so the interpretation of the mean and standard deviation should be taken with that in mind.  For instance, the standard deviation can exceed the mean and/or median (e.g., $T{\scriptstyle_{\alpha, tot}}$).  The summary of the observations are as follows:

\begin{itemize}[itemsep=0pt,parsep=0pt,topsep=0pt]
  \item[]  \textbf{Scalar Temperatures}
  \begin{itemize}[itemsep=0pt,parsep=0pt,topsep=0pt]
    \item  $T{\scriptstyle_{e, tot}}$ $=$ 12.2(11.9)[3.2] eV
    \item  $T{\scriptstyle_{p, tot}}$ $=$ 12.7(8.6)[14.1] eV
    \item  $T{\scriptstyle_{\alpha, tot}}$ $=$ 23.9(10.8)[31.7] eV
  \end{itemize}
  \item[]  \textbf{Temperature Ratios}
  \begin{itemize}[itemsep=0pt,parsep=0pt,topsep=0pt]
    \item  $\left(T{\scriptstyle_{e}}/T{\scriptstyle_{p}}\right){\scriptstyle_{tot}}$ $=$ 1.64(1.27)[1.26]
    \item  $\left(T{\scriptstyle_{e}}/T{\scriptstyle_{\alpha}}\right){\scriptstyle_{tot}}$ $=$ 1.24(0.82)[1.25]
    \item  $\left(T{\scriptstyle_{\alpha}}/T{\scriptstyle_{p}}\right){\scriptstyle_{tot}}$ $=$ 2.50(1.94)[1.45]
  \end{itemize}
  \item[]  \textbf{Plasma Betas}
  \begin{itemize}[itemsep=0pt,parsep=0pt,topsep=0pt]
    \item  $\beta{\scriptstyle_{e, tot}}$ $=$ 2.31(1.09)[17.6]
    \item  $\beta{\scriptstyle_{p, tot}}$ $=$ 1.79(1.05)[11.4]
    \item  $\beta{\scriptstyle_{\alpha, tot}}$ $=$ 0.17(0.05)[1.35].
  \end{itemize}
  \item[]  \textbf{Collision Rates}
  \begin{itemize}[itemsep=0pt,parsep=0pt,topsep=0pt]
    \item $\nu{\scriptstyle_{ee}}$ $\sim$ $4 \times 10^{-6}$($2 \times 10^{-6}$)[$4 \times 10^{-6}$] \# $s^{-1}$
    \item $\nu{\scriptstyle_{pp}}$ $\sim$ $3 \times 10^{-7}$($1 \times 10^{-7}$)[$4 \times 10^{-7}$] \# $s^{-1}$
    \item $\nu{\scriptstyle_{\alpha \alpha}}$ $\sim$ $6 \times 10^{-8}$($2 \times 10^{-8}$)[$1 \times 10^{-7}$] \# $s^{-1}$
    \item $\nu{\scriptstyle_{ep}}$ $\sim$ $5 \times 10^{-6}$($3 \times 10^{-6}$)[$4 \times 10^{-6}$] \# $s^{-1}$
    \item $\nu{\scriptstyle_{e \alpha}}$ $\sim$ $7 \times 10^{-7}$($4 \times 10^{-7}$)[$1 \times 10^{-6}$] \# $s^{-1}$
    \item $\nu{\scriptstyle_{p \alpha}}$ $\sim$ $3 \times 10^{-8}$($1 \times 10^{-8}$)[$7 \times 10^{-8}$] \# $s^{-1}$
    \item $\nu{\scriptstyle_{iaw}}$ $\sim$ $5 \times 10^{-4}$($5 \times 10^{-4}$)[$2 \times 10^{-4}$] \# $s^{-1}$ (for 0.1 mV/m wave amplitude)
  \end{itemize}
\end{itemize}

\indent  The results are relevant for long-term statistical models and parameter range limits in empirical models.  These observations are also relevant to comparisons with astrophysical plasmas like the intra-galaxy-cluster medium.  Finally, this work will provide a statistical baseline for the upcoming \emph{Solar Orbiter} and \emph{Parker Solar Probe} missions and future IMAP mission.

\acknowledgments
\noindent  The authors thank A.F.- Vi{\~n}as and T. Nieves-Chinchilla for useful discussions of solar wind plasma physics.  L.B.W. was partially supported by \emph{Wind} MO\&DA grants.  C.S.S. was partially supported by NASA grant NNX16AI59G and NSF SHINE grant 1622498.  S.D.B., T.A.B., C.S.S., and M.P.P. were partially supported by NASA grant NNX16AP95G.  B.A.M. was partially supported by NASA grants NNX17AC72G and NNX17AH88G.  K.G.K. and J.C.K. were partially supported by NASA grant NNX14AR78G.  M.L.S. was partially supported by grants NNX14AT26G
and NNX13AI75G.  The CFA Interplanetary Shock Database is supported by NASA grant NNX13AI75G.  The authors thank the Harvard Smithsonian Center for Astrophysics, the NASA SPDF/CDAWeb team, and the \emph{Wind} team for the interplanetary shock analysis, \emph{Wind} plasma and magnetic field data, and the \emph{Wind} ICME catalog, respectively.  The \emph{Wind} shock database can be found at: \\
\noindent  \url{https://www.cfa.harvard.edu/shocks/wi\_data/}.  \\
\noindent  The \emph{Wind} ICME catalog can be found at: \\
\noindent  \url{https://wind.nasa.gov/ICMEindex.php}.  \\
\noindent  Analysis software used herein can be found at: \\
\noindent  \url{https://github.com/lynnbwilsoniii/wind\_3dp\_pros}.  \\

\appendix
\section{Definitions}  \label{app:Definitions}

\indent  In this appendix we summarize the symbols and parameters we use throughout.  In the following, for all temperature-dependent parameters we define $j$ to represent the entire distribution ($j$ $=$ $tot$), the parallel component ($j$ $=$ $\parallel$), and the perpendicular component ($j$ $=$ $\perp$).  The parallel and perpendicular are with respect to the quasi-static magnetic field vector, $\textbf{\textit{B}}{\scriptstyle_{o}}$ [nT].

\indent  The standard one-variable statistics symbols are defined as follows:  minimum $\equiv$ $X{\scriptstyle_{min}}$, maximum $\equiv$ $X{\scriptstyle_{max}}$, mean $\equiv$ $\bar{X}$, median $\equiv$ $\tilde{X}$, standard deviation $\equiv$ $\sigma{\scriptstyle_{x}}$, lower quartile $\equiv$ $X{\scriptstyle_{25\%}}$, and upper quartile $\equiv$ $X{\scriptstyle_{75\%}}$.  We use the following parameter definitions herein:  $\varepsilon{\scriptstyle_{o}}$ and $\mu{\scriptstyle_{o}}$ are the permittivity and permeability of free space; $k{\scriptstyle_{B}}$ is the Boltzmann constant; $n{\scriptstyle_{s}}$ is the number density of species $s$ [$cm^{-3}$]; $m{\scriptstyle_{s}}$ is the mass of species $s$ [$kg$]; $q{\scriptstyle_{s}}$ $=$ $Z{\scriptstyle_{s}} \ e$ is the charge of species $s$, where $Z{\scriptstyle_{s}}$ and $e$ are the charge state and fundamental charge, respectively [$C$]; $\textbf{\textit{V}}{\scriptstyle_{s}}$ is the bulk flow velocity of species $s$ [$km \ s^{-1}$]; and $\nu{\scriptstyle_{ss'}}$ is the classical binary Coulomb collision frequency \citep[e.g.,][]{krall73a, spitzer53a} for a 90$^{\circ}$ deflection angle between particles of species $s$ and $s'$ [\# $s^{-1}$].

\indent  Further, we define the reduced mass ($\mu{\scriptstyle_{ss'}}$) [$kg$], scalar temperature ($T{\scriptstyle_{s, tot}}$) [$eV$], most probable thermal speed of a one-dimensional velocity distribution ($V{\scriptstyle_{Ts, j}}$) [$km \ s^{-1}$], plasma frequency ($\omega{\scriptstyle_{ps}}$) [$rad \ s^{-1}$], plasma beta ($\beta{\scriptstyle_{s, j}}$) [N/A], j$^{th}$ component temperature ratio ($\left(T{\scriptstyle_{s'}}/T{\scriptstyle_{s}}\right){\scriptstyle_{j}}$) [N/A], and electron Debye length ($\lambda{\scriptstyle_{De}}$) [$m$] of species $s$ and $s'$ as:

\begin{subequations}
  \begin{align}
    \mu{\scriptstyle_{ss'}} & = \frac{ m{\scriptstyle_{s}} \ m{\scriptstyle_{s'}} }{ \left( m{\scriptstyle_{s}} + m{\scriptstyle_{s'}} \right) } \label{eq:params_0} \\
    T{\scriptstyle_{s, tot}} & = \frac{1}{3} \left( T{\scriptstyle_{s, \parallel}} + 2 \ T{\scriptstyle_{s, \perp}} \right) \label{eq:params_1} \\
    V{\scriptstyle_{Ts, j}} & = \sqrt{ \frac{ 2 \ k{\scriptstyle_{B}} \ T{\scriptstyle_{s, j}} }{ m{\scriptstyle_{s}} } } \label{eq:params_2} \\
    \omega{\scriptstyle_{ps}} & = \sqrt{ \frac{ n{\scriptstyle_{s}} \ q{\scriptstyle_{s}}^{2} }{ \varepsilon{\scriptstyle_{o}} \ m{\scriptstyle_{s}} } } \label{eq:params_3} \\
    \beta{\scriptstyle_{s, j}} & = \frac{ 2 \mu{\scriptstyle_{o}} n{\scriptstyle_{s}} k{\scriptstyle_{B}} T{\scriptstyle_{s, j}} }{ \lvert \textbf{\textit{B}}{\scriptstyle_{o}} \rvert^{2} } \label{eq:params_4} \\
    \left( \frac{ T{\scriptstyle_{s'}} }{ T{\scriptstyle_{s}} } \right){\scriptstyle_{j}} & = \left( \frac{ T{\scriptstyle_{s', j}} }{ T{\scriptstyle_{s, j}} } \right) \label{eq:params_5} \\
    \lambda{\scriptstyle_{De}} & = \frac{ V{\scriptstyle_{Te, tot}} }{ \sqrt{ 2 } \ \omega{\scriptstyle_{pe}} } = \sqrt{ \frac{ \varepsilon{\scriptstyle_{o}} \ k{\scriptstyle_{B}} \ T{\scriptstyle_{e, tot}} }{ n{\scriptstyle_{e}} \ e^{2} } } \label{eq:params_6}
  \end{align}
\end{subequations}

\noindent  These definitions are used throughout.

\section{Coulomb Collisions}  \label{app:CoulombCollisions}

\indent  The various forms of the Coulomb collision rates \citep[e.g.,][]{hernandez85a, hinton84a, krall73a, schunk75a, schunk77a, spitzer53a}, $\nu{\scriptstyle_{ss'}}$, relevant to this work can be approximated as:

\begin{subequations}
  \begin{align}
    \nu{\scriptstyle_{ee}} & \simeq \frac{ 4 \sqrt{ \pi } \ n{\scriptstyle_{e}} \ e^{4} }{ 3 \left( 4 \pi \varepsilon{\scriptstyle_{o}} \right)^{2} \ m{\scriptstyle_{e}}^{2} \ V{\scriptstyle_{Te, tot}}^{3} } \ \ln{\Lambda{\scriptstyle_{ee}}}  \label{eq:SHcollfreq_0} \\
    \nu{\scriptstyle_{pp}} & \simeq \frac{ 4 \sqrt{ \pi } \ n{\scriptstyle_{p}} \ e^{4} }{ 3 \left( 4 \pi \varepsilon{\scriptstyle_{o}} \right)^{2} \ m{\scriptstyle_{p}}^{2} \ V{\scriptstyle_{Tp, tot}}^{3} } \ \ln{\Lambda{\scriptstyle_{pp}}}  \label{eq:SHcollfreq_1} \\
    \nu{\scriptstyle_{\alpha \alpha}} & \simeq \frac{ 64 \sqrt{ \pi } \ n{\scriptstyle_{\alpha}} \ e^{4} }{ 3 \left( 4 \pi \varepsilon{\scriptstyle_{o}} \right)^{2} \ m{\scriptstyle_{\alpha}}^{2} \ V{\scriptstyle_{T \alpha, tot}}^{3} } \ \ln{\Lambda{\scriptstyle_{\alpha \alpha}}}  \label{eq:SHcollfreq_2} \\
    \nu{\scriptstyle_{ep}} & \simeq \frac{ 2 \sqrt{ 4 \pi } \ n{\scriptstyle_{p}} \ e^{4} }{ 3 \left( 4 \pi \varepsilon{\scriptstyle_{o}} \right)^{2} \ \mu{\scriptstyle_{ep}}^{2} \ V{\scriptstyle_{Tep}}^{3} } \ \ln{\Lambda{\scriptstyle_{ep}}}  \label{eq:SHcollfreq_3} \\
    \nu{\scriptstyle_{e \alpha}} & \simeq \frac{ 8 \sqrt{ 4 \pi } \ n{\scriptstyle_{\alpha}} \ e^{4} }{ 3 \left( 4 \pi \varepsilon{\scriptstyle_{o}} \right)^{2} \ \mu{\scriptstyle_{e \alpha}}^{2} \ V{\scriptstyle_{Te \alpha}}^{3} } \ \ln{\Lambda{\scriptstyle_{e \alpha}}}  \label{eq:SHcollfreq_4} \\
    \nu{\scriptstyle_{p \alpha}} & \simeq \frac{ 8 \sqrt{ 2 \pi } \ n{\scriptstyle_{\alpha}} \ e^{4} }{ 3 \left( 4 \pi \varepsilon{\scriptstyle_{o}} \right)^{2} \ \mu{\scriptstyle_{p \alpha}}^{2} \ V{\scriptstyle_{Tp \alpha}}^{3} } \ \ln{\Lambda{\scriptstyle_{p \alpha}}}  \label{eq:SHcollfreq_5} \\
  \end{align}
\end{subequations}

\noindent  where $V{\scriptstyle_{Tss'}}^{2}$ $=$ $V{\scriptstyle_{Ts, tot}}^{2} + V{\scriptstyle_{Ts', tot}}^{2}$ and we have approximated the $\Lambda{\scriptstyle_{ss'}}$ terms in the Coulomb logarithm as:

\begin{equation}
  \label{eq:SHcollfreq_6}
  \Lambda{\scriptstyle_{ss'}} \simeq \frac{ \left( 4 \pi \varepsilon{\scriptstyle_{o}} \right) \ \mu{\scriptstyle_{ss'}} \ V{\scriptstyle_{Tss'}}^{2} }{ \sqrt{2} \ Z{\scriptstyle_{s}} \ Z{\scriptstyle_{s'}} \ e^{2} } \left[ \left( \frac{ \omega{\scriptstyle_{ps}} }{ V{\scriptstyle_{Ts, tot}} } \right)^{2} + \left( \frac{ \omega{\scriptstyle_{ps'}} }{ V{\scriptstyle_{Ts', tot}} } \right)^{2} \right]^{-1/2}
\end{equation}

\noindent  Note that the rates in Equations \ref{eq:SHcollfreq_0}--\ref{eq:SHcollfreq_5} use the effective thermal speed given by $V{\scriptstyle_{Tss'}}$ instead of the more commonly used single-species value \citep[e.g.,][]{hernandez85a}.  This makes all forms of $\nu{\scriptstyle_{ss'}}$ smaller and does not significantly affect most of the electron-dependent rates since $V{\scriptstyle_{Te}}$ $\gg$ $V{\scriptstyle_{Tp}}$ and $V{\scriptstyle_{T \alpha}}$.

\indent  The quasi-linear approximation for the effective collision rates between ion-acoustic waves and particles \citep[e.g., see][and references therein]{wilsoniii14a, wilsoniii14b} is given by:
\begin{equation}
  \label{eq:WPIcollfreq_0}
  \nu{\scriptstyle_{iaw}} = \omega{\scriptstyle_{pe}} \ \frac{ \varepsilon{\scriptstyle_{o}} \ \lvert \delta E \rvert^{2} }{ 2 \ n{\scriptstyle_{e}} \ k{\scriptstyle_{B}} \ T{\scriptstyle_{e, tot}} }
\end{equation}

\noindent  where $\delta E$ is the wave amplitude.

\clearpage
\section{Previous Measurements}  \label{app:PreviousMeasurements}

\indent  In this appendix, we summarize, by way of tables, the  observations of previous solar wind studies of the parameters examined herein.

\startlongtable  
\begin{deluxetable*}{| l | c | c | c | c | c | c |}
  \tabletypesize{\footnotesize}    
  \tablecaption{Measurements of Solar Wind Temperatures at 1 AU \label{tab:PrevTemperatures}}
  \tablehead{\colhead{Reference} & \colhead{Temperature [eV]} & \colhead{Spacecraft} & \colhead{Notes} & \colhead{Range} & \colhead{$\bar{\mathbf{X}}$}\tablenotemark{c} & \colhead{$\tilde{\mathbf{X}}$}\tablenotemark{d}}
  \startdata
  \citet[][]{lepri13a} & $T{\scriptstyle_{p, tot}}$ & \multirow{3}{*}{ACE} & \multirow{3}{*}{SW\tablenotemark{a}} & $\sim$0.3--216 & & \\
  \multirow{2}{*}{\citet[][]{skoug00a}} & $T{\scriptstyle_{e, tot}}$ & & & $\sim$3.4--78 & $\sim$14.8 & \\
                       & $T{\scriptstyle_{p, tot}}$ & & & $\sim$0.2--87 & $\sim$7.35 & \\
  \hline
  \citet[][]{serbu72a} & $T{\scriptstyle_{e, tot}}$ & \multirow{2}{*}{\emph{Explorer} 34} & \multirow{2}{*}{SW} & $\sim$4.3--35 & &  \\
  \citet[][]{hundhausen70b} & $T{\scriptstyle_{p, tot}}$ & & & $<$ 1 to $>$25 & $\sim$4 &  \\
  \hline
  \multirow{2}{*}{\citet[][]{reisenfeld13a}} & $T{\scriptstyle_{\alpha, tot}}$ & \multirow{2}{*}{\emph{Genesis}} & \multirow{2}{*}{SW} & $\sim$0.1--345 &  &  \\
  & $T{\scriptstyle_{p, tot}}$ & & & & 10.9$\pm$8.1 &  \\
  \hline
  \multirow{2}{*}{\citet[][]{marsch82b}} & $T{\scriptstyle_{\alpha, \parallel}}$ & \multirow{18}{*}{\emph{Helios} 1 \& 2} & \multirow{11}{*}{SW} & $\sim$10--78 &  &  \\
  & $T{\scriptstyle_{\alpha, \perp}}$ & & & $\sim$8--60 &  &  \\
  \multirow{2}{*}{\citet[][]{marsch82c}} & $T{\scriptstyle_{p, \parallel}}$ & & & $\sim$4--32 &  &  \\
  & $T{\scriptstyle_{p, \perp}}$ & & & $\sim$3--23 &  &  \\
  \multirow{2}{*}{\citet[][]{marsch89a}} & $T{\scriptstyle_{e, tot}}$ & & & $\sim$8.5--17 &  &  \\
  & $T{\scriptstyle_{p, tot}}$ & & & $\sim$4.3--28 &  &  \\
  \multirow{3}{*}{\citet[][]{pilipp87c}} & $T{\scriptstyle_{e, \parallel}}$ & & & $\sim$6--26 &  &  \\
  & $T{\scriptstyle_{e, \perp}}$ & & & $\sim$6--26 &  &  \\
  & $T{\scriptstyle_{p, tot}}$ & & & $\sim$1--80 &  &  \\
  \multirow{9}{*}{\citet[][]{schwenn90a}} & \multirow{3}{*}{$T{\scriptstyle_{p, tot}}$} & & Slow\tablenotemark{b} SW & & $\sim$2.9--4.7 &  \\
  & & & Fast SW & & $\sim$20--25 &  \\
  & & &  All SW & & $\sim$10 &  \\
  & \multirow{3}{*}{$T{\scriptstyle_{e, tot}}$} & & Slow SW & & $\sim$11--17 &  \\
  & & & Fast SW & & $\sim$8.6--11 &  \\
  & & &  All SW & & $\sim$12 &  \\
  & \multirow{3}{*}{$T{\scriptstyle_{\alpha, tot}}$} & & Slow SW & & $\sim$9.5--15 &  \\
  & & & Fast SW & & $\sim$62--122  &  \\
  & & &  All SW & & $\sim$50  &  \\
  \hline
  \citet[][]{feldman73c} & $T{\scriptstyle_{p, tot}}$ & \multirow{3}{*}{\emph{Imp} 6, 7, \& 8} & \multirow{3}{*}{SW} & $<$ 1 to $>$25 & &  \\
  \citet[][]{feldman78a} & $T{\scriptstyle_{p, tot}}$ & & & $<$ 2.5 to $>$25 & &  \\
  \citet[][]{feldman79b} & $T{\scriptstyle_{e, tot}}$ & & & $<$ 8.5 to $>$18 & &  \\
  \hline
  \multirow{2}{*}{\citet[][]{mccomas89a}} & $T{\scriptstyle_{e, tot}}$ & \multirow{8}{*}{ISEE 3} & \multirow{2}{*}{SW} & $\sim$4.3--26 &  &  \\
  & $T{\scriptstyle_{p, tot}}$ & & & $\sim$0.9--26 &  &  \\
  \multirow{6}{*}{\citet[][]{newbury98a}} & \multirow{3}{*}{$T{\scriptstyle_{e, tot}}$} &  & SW & $\sim$4.3--38 & $\sim$12.2 &  \\
  & & & Slow SW & $\sim$4.3--32 & $\sim$11.1 &  \\
  & & & Fast SW & $\sim$4.3--31 & $\sim$12.2 &  \\
  & \multirow{3}{*}{$T{\scriptstyle_{p, tot}}$} & & SW & $\sim$1--69 &  &  \\
  & & & Slow SW & $\sim$1--26 & $\sim$3.9 &  \\
  & & & Fast SW & $\sim$1--69 & $\sim$15.3 &  \\
  \citet[][]{phillips89b} & $T{\scriptstyle_{e, tot}}$ & & SW & $\sim$8.5--21 & &  \\
  \hline
  \multirow{2}{*}{\citet[][]{elliott12a}} & \multirow{4}{*}{$T{\scriptstyle_{p, tot}}$} & \multirow{2}{*}{All OMNI} & SW & $\sim$0.2--276 & &  \\
  & & & No ICMEs & $\sim$0.5--233 & &  \\
  \citet[][]{elliott16a} & & Late OMNI & & $\sim$0.6--86 & &  \\
  \hline
  \citet[][]{jian14a} & $T{\scriptstyle_{p, tot}}$ & STEREO & SW & $<$ 1 to $>$30 & & $\sim$6 \\
  \hline
  \citet[][]{gosling72a} & \multirow{2}{*}{$T{\scriptstyle_{p, tot}}$} & \multirow{4}{*}{\emph{Vela} 3 \& 4} & \multirow{4}{*}{SW} & $\sim$4--35 & &  \\
  \citet[][]{hundhausen70b} & & & & $<$ 1 to $>$17 & $\sim$8 & $\sim$6  \\
  \multirow{2}{*}{\citet[][]{montgomery68a}} & $T{\scriptstyle_{e, tot}}$ & & & $\sim$6--17 &  &  \\
  & $T{\scriptstyle_{p, tot}}$ & & & $\sim$2--28 &  &  \\
  \hline
  \citet[][]{maksimovic05a} & $T{\scriptstyle_{e, tot}}$ & \multirow{6}{*}{\emph{Wind}} & \multirow{6}{*}{SW} & $\sim$6.5--17 & & \\
  \multirow{3}{*}{\citet[][]{maruca11a}} & $T{\scriptstyle_{p, tot}}$ & & & $\sim$2.5--35 &  &  \\
  & $T{\scriptstyle_{p, \parallel}}$ & & & $\sim$1.7--50 &  &  \\
  & $T{\scriptstyle_{p, \perp}}$ & & & $\sim$2.5--53 &  &  \\
  \citet[][]{salem01a} & $T{\scriptstyle_{e, tot}}$ & & & $\sim$10--26 &  &  \\
  \citet[][]{salem03a} & $T{\scriptstyle_{e, tot}}$ & & & $\sim$5--23 &  &  \\
  \enddata
  \tablenotetext{a}{SW $\equiv$ Solar Wind, a generic term for ambient/all solar wind conditions}
  \tablenotetext{b}{Fast and Slow SW are typically defined as bulk flow speed above or below, respectively, some threshold (typically $\sim$350--500 $km \ s^{-1}$)}
  \tablenotetext{c}{mean or average}
  \tablenotetext{d}{median}
  \tablecomments{OMNI is a dataset comprised of multiple spacecraft from SPDF/CDAWeb, where All refers to 1963--Present and Late to 1978--Present.  For symbol definitions, see Appendix \ref{app:Definitions}.}
\end{deluxetable*}

\startlongtable  
\begin{deluxetable*}{| l | c | c | c | c | c | c |}
  \tabletypesize{\small}    
  \tablecaption{Measurements of Solar Wind Temperature Ratios at 1 AU \label{tab:PrevTempRatios}}
  \tablehead{\colhead{Reference} & \colhead{Ratios [N/A]} & \colhead{Spacecraft} & \colhead{Notes} & \colhead{Range} & \colhead{$\bar{\mathbf{X}}$} & \colhead{$\tilde{\mathbf{X}}$}}
  \startdata
  \multirow{5}{*}{\citet[][]{skoug00a}} & \multirow{5}{*}{$\left(T{\scriptstyle_{e}}/T{\scriptstyle_{p}}\right){\scriptstyle_{tot}}$} & \multirow{7}{*}{ACE} & SW & & $\sim$3.46 &  \\
  & & & NCS\tablenotemark{a} & & $\sim$3.25 &  \\
  & & & CS\tablenotemark{b} & & $\sim$4.19 &  \\
  & & & MC\tablenotemark{c} & & $\sim$6.80 &  \\
  & & & non-MC CS & & $\sim$3.25 &  \\
  \citet[][]{tracy15a} & \multirow{2}{*}{$\left(T{\scriptstyle_{\alpha}}/T{\scriptstyle_{p}}\right){\scriptstyle_{tot}}$} & & \multirow{2}{*}{SW} & $\sim$0.16 to $>$9 & &  \\
  \citet[][]{tracy16a} & & & & & & $\sim$1.78--6.92  \\
  \hline
  \citet[][]{sahraoui13a} & $\left(T{\scriptstyle_{e}}/T{\scriptstyle_{i}}\right){\scriptstyle_{tot}}$ & \emph{Cluster} & SW & $<$ 0.1 to $>$10 &  &  \\
  \hline
  \multirow{2}{*}{\citet[][]{burlaga70a}} & $\left(T{\scriptstyle_{e}}/T{\scriptstyle_{p}}\right){\scriptstyle_{tot}}$ & \multirow{2}{*}{\emph{Explorer} 34} & \multirow{2}{*}{SW} & $\sim$1.5--5.0 &  &  \\
  & $\left(T{\scriptstyle_{\alpha}}/T{\scriptstyle_{p}}\right){\scriptstyle_{tot}}$ & & & $\sim$3.2--4.2 & $\sim$3.75 &  \\
  \hline
  \citet[][]{marsch82b} & \multirow{3}{*}{$\left(T{\scriptstyle_{\alpha}}/T{\scriptstyle_{p}}\right){\scriptstyle_{tot}}$} & \multirow{3}{*}{\emph{Helios} 1 \& 2} & SW & $\sim$2.2--3.1 &  &  \\
  \multirow{2}{*}{\citet[][]{schwenn90a}} & & & Slow SW & & $\sim$2.9 &  \\
  & & & Fast SW & & $\sim$3.0 &  \\
  \hline
  \citet[][]{feldman74a} & \multirow{2}{*}{$\left(T{\scriptstyle_{\alpha}}/T{\scriptstyle_{p}}\right){\scriptstyle_{tot}}$} & \multirow{6}{*}{\emph{Imp} 6, 7, \& 8} & \multirow{4}{*}{SW} & $\sim$1--6 & &  \\
  \citet[][]{feldman78a} & & & & $<$ 2.5 to $>$6.5 &  &  \\
  \citet[][]{feldman79b} & $\left(T{\scriptstyle_{e}}/T{\scriptstyle_{p}}\right){\scriptstyle_{tot}}$ & & & $<$ 0.5 to $>$5 &  &  \\
  \multirow{3}{*}{\citet[][]{schwenn90a}} & \multirow{3}{*}{$\left(T{\scriptstyle_{\alpha}}/T{\scriptstyle_{p}}\right){\scriptstyle_{tot}}$} & & &  & $\sim$4.9 &  \\
  & & & Slow SW & & $\sim$3.2 &  \\
  & & & Fast SW & & $\sim$6.2 &  \\
  \hline
  \multirow{2}{*}{\citet[][]{richardson97a}} & \multirow{2}{*}{$\left(T{\scriptstyle_{e}}/T{\scriptstyle_{p}}\right){\scriptstyle_{tot}}$} & \multirow{6}{*}{ISEE 3} & MC & $>$1 (All) & $\sim$4.7 &  \\
  & & & ICMEs & $>$1 (67/68)\tablenotemark{e} & $\sim$3.7 &  \\
  \multirow{4}{*}{\citet[][]{newbury98a}} & \multirow{4}{*}{$\left(T{\scriptstyle_{e}}/T{\scriptstyle_{p}}\right){\scriptstyle_{tot}}$} & & SW & $\sim$0.1--8.0 & $\sim$2.3 &  \\
  & & & Slow SW & & $\sim$3.7 &  \\
  & & & Int. SW & & $\sim$2.0 &  \\
  & & & Fast SW & & $\sim$1.1 &  \\
  \hline
  \citet[][]{neugebauer76a} & $\left(T{\scriptstyle_{\alpha}}/T{\scriptstyle_{p}}\right){\scriptstyle_{tot}}$ & \emph{Ogo} 5 & SW & $\sim$1.5--9.5 &  &  \\
  \hline
  \citet[][]{jian14a} & $\left(T{\scriptstyle_{\alpha}}/T{\scriptstyle_{p}}\right){\scriptstyle_{tot}}$ & STEREO & SW & $\sim$2.5--20 &  & $\sim$7.5 \\
  \hline
  \citet[][]{montgomery68a} & $\left(T{\scriptstyle_{e}}/T{\scriptstyle_{p}}\right){\scriptstyle_{tot}}$ & \multirow{4}{*}{\emph{Vela} 3 \& 4} & \multirow{3}{*}{SW} & $\sim$1.5--5.0 &  &  \\
  \citet[][]{robbins70a} & \multirow{2}{*}{$\left(T{\scriptstyle_{\alpha}}/T{\scriptstyle_{p}}\right){\scriptstyle_{tot}}$} & & & $\sim$0.2--10 &  &  \\
  \citet[][]{hirshberg74a} & & & & $\sim$2.5 to $>$7 &  &  \\
  \multirow{2}{*}{\citet[][]{hundhausen70b}} & \multirow{2}{*}{$\left(T{\scriptstyle_{e}}/T{\scriptstyle_{p}}\right){\scriptstyle_{tot}}$} & & Slow SW & $\sim$0.4 to $>$10 &  &  \\
  & & & Fast SW & $\sim$0.2--6.0 &  &  \\
  \hline
  \multirow{3}{*}{\citet[][]{kasper08a}} & $\left(T{\scriptstyle_{\alpha}}/T{\scriptstyle_{p}}\right){\scriptstyle_{tot}}$ & \multirow{7}{*}{\emph{Wind}} & \multirow{7}{*}{SW} & $<$ 0.75 to $>$8 &  &  \\
  & $\left(T{\scriptstyle_{\alpha}}/T{\scriptstyle_{p}}\right){\scriptstyle_{\perp}}$ & & & $\sim$1--6 &  &  \\
  & $\left(T{\scriptstyle_{\alpha}}/T{\scriptstyle_{p}}\right){\scriptstyle_{\parallel}}$ & & & $\sim$3.0--5.5 &  &  \\
  \citet[][]{kasper13a} & $\left(T{\scriptstyle_{\alpha}}/T{\scriptstyle_{p}}\right){\scriptstyle_{\perp}}$ & & & $<$ 4 to $>$7 &  &  \\
  \citet[][]{kasper17b} & $\left(T{\scriptstyle_{\alpha}}/T{\scriptstyle_{p}}\right){\scriptstyle_{tot}}$ & & & $<$ 1 to $>$8 &  &  \\
  \citet[][]{maruca13b} & $\left(T{\scriptstyle_{\alpha}}/T{\scriptstyle_{p}}\right){\scriptstyle_{tot}}$ & & & $<$ 0.75 to $>$10 &  &  \\
  \citet[][]{vech17a} & $\left(T{\scriptstyle_{e}}/T{\scriptstyle_{p}}\right){\scriptstyle_{tot}}$ & & & $\sim$0.3--1.6 &  &  \\
  \hline
  \enddata
  \tablenotetext{a}{no counter-streaming electrons present}
  \tablenotetext{b}{counter-streaming electrons present}
  \tablenotetext{c}{magnetic cloud}
  \tablenotetext{e}{number of ICMEs satisfying this condition}
  \tablecomments{Definitions/Symbols are the same as in Table \ref{tab:PrevTemperatures}.  For symbol definitions, see Appendix \ref{app:Definitions}.}
\end{deluxetable*}

\begin{deluxetable*}{| l | c | c | c | c | c | c |}
\startlongtable  
  \tabletypesize{\small}    
  \tablecaption{Measurements of Solar Wind betas at 1 AU \label{tab:PrevBetas}}
  \tablehead{\colhead{Reference} & \colhead{Betas [N/A]} & \colhead{Spacecraft} & \colhead{Notes} & \colhead{Range} & \colhead{$\bar{\mathbf{X}}$} & \colhead{$\tilde{\mathbf{X}}$}}
  \startdata
  \citet[][]{sahraoui13a} & $\beta{\scriptstyle_{i, tot}}$ & \emph{Cluster} & SW & $<$ 0.1 to $\sim$5 & &  \\
  \hline
  \citet[][]{reisenfeld13a} & $\beta{\scriptstyle_{i, tot}}$ & \emph{Genesis} & SW & & 0.53$\pm$0.60 &  \\
  \hline
  \citet[][]{matteini07a} & $\beta{\scriptstyle_{p, \parallel}}$ & \emph{Helios} 1 \& 2 & SW & $\sim$0.16--4.0 &  &  \\
  \hline
  \multirow{2}{*}{\citet[][]{ness71a}} & \multirow{2}{*}{$\beta{\scriptstyle_{p, tot}}$} & \multirow{2}{*}{\emph{Imp} 3 \& \emph{Vela} 3} & SW & $\sim$0.09--2.5 & $\sim$0.95 & $\sim$0.81 \\
  & & & Slow SW & & $\sim$0.78 & $\sim$0.67 \\
  \hline
  \citet[][]{reisenfeld13a} & $\beta{\scriptstyle_{i, tot}}$ & All OMNI & SW & & 0.75$\pm$0.99 &  \\
  \hline
  \citet[][]{jian14a} & $\beta{\scriptstyle_{p, tot}}$ & STEREO & SW & $<$ 1 to $>$100 & & $\sim$1.55 \\
  \hline
  \multirow{2}{*}{\citet[][]{adrian16a}} & $\beta{\scriptstyle_{e, \parallel}}$ & \multirow{12}{*}{\emph{Wind}} & Slow SW & $\sim$0.02 to $>$10 & & \\
  & $\beta{\scriptstyle_{e, \perp}}$ & & Fast SW & $\sim$0.05 to $>$10 & & \\
  \citet[][]{bale09a} & $\beta{\scriptstyle_{p, \parallel}}$ & & \multirow{10}{*}{SW} & $<$ 0.001 to $>$100 & & \\
  \citet[][]{bale13a} & $\beta{\scriptstyle_{e, tot}}$ & & & $<$ 0.01 to $>$100 & & \\
  \citet[][]{hellinger06a} & \multirow{7}{*}{$\beta{\scriptstyle_{p, \parallel}}$} & & & $<$ 0.002 to $>$30 & & \\
  \citet[][]{hellinger14a} & & & & $<$ 0.001 to $>$30 & & \\
  \citet[][]{kasper06a}    & & & & $<$ 0.2 to $>$100 & & \\
  \citet[][]{kasper13a}    & & & & $<$ 0.08 to $>$15 & & \\
  \citet[][]{maruca11a}    & & & & $<$ 0.03 to $>$30 & & \\
  \citet[][]{maruca12a}    & & & & $\sim$0.02 to $>$30 & & \\
  \citet[][]{vech17a}      & & & & $\sim$0.01--100 & & \\
  \citet[][]{maruca12a}    & $\beta{\scriptstyle_{\alpha, \parallel}}$ & & & $<$ 0.003 to $\sim$6 & & \\
  \hline
  \enddata
  \tablecomments{Definitions/Symbols are the same as in Table \ref{tab:PrevTemperatures}.  For symbol definitions, see Appendix \ref{app:Definitions}.}
\end{deluxetable*}

\clearpage

\end{document}